\documentclass[conference]{IEEEtran}
\usepackage{xspace}
\usepackage{balance}
\usepackage{graphicx}
\usepackage[hyphens]{url}
\usepackage{epstopdf}
\usepackage{microtype}
\usepackage{booktabs}
\usepackage[table,xcdraw]{xcolor}
\usepackage{caption}
\usepackage{subcaption}
\usepackage{amsmath}
\usepackage{enumitem}

\newcommand{\fakepar}[1]{\smallbreak\noindent{}}
\newcommand{\boldpar}[1]{\smallbreak\noindent\textbf{#1.}}

\newcommand{\ieee}{\mbox{IEEE~802.15.4}\xspace}
\newcommand{\wifi}{\mbox{Wi-Fi}\xspace}
\newcommand{\blefive}{\mbox{BLE\,5}\xspace}
\newcommand{\dcubee}{\mbox{D-Cube}\xspace}
\newcommand{\jamlabng}{\mbox{JamLab-NG}\xspace}

\iftrue 
	\newcommand{\cb}[1]{\footnote{{\color{red}\bf Carlo: #1}\color{black}}}
	\newcommand{\ms}[1]{\footnote{{\color{red}\bf Markus: #1}\color{black}}}
	\newcommand{\mb}[1]{\footnote{{\color{red}\bf Michael: #1}\color{black}}}
	\newcommand{\aem}[1]{\footnote{{\color{red}\bf Antonio: #1}\color{black}}}
	\newcommand{\xm}[1]{\footnote{{\color{red}\bf Xiaoyuan: #1}\color{black}}}
	\newcommand{\yl}[1]{\footnote{{\color{red}\bf Ye: #1}\color{black}}}
	\newcommand{\ur}[1]{\footnote{{\color{red}\bf Usman: #1}\color{black}}}
	\newcommand{\as}[1]{\footnote{{\color{red}\bf Aleksandar: #1}\color{black}}}
	\newcommand{\kr}[1]{\footnote{{\color{red}\bf Kay: #1}\color{black}}}
\else
	\newcommand{\cb}[1]{}
	\newcommand{\ms}[1]{}
	\newcommand{\mb}[1]{}
	\newcommand{\aem}[1]{}
	\newcommand{\xm}[1]{}
	\newcommand{\yl}[1]{}
	\newcommand{\ur}[1]{}
	\newcommand{\as}[1]{}
	\newcommand{\kr}[1]{}
\fi

\begin{document}

% Title
\title{
The Impact of the Physical Layer on the\\ Performance of Concurrent Transmissions \vspace{-1.50mm} \\
}

% Authors (blinded --> iftrue, unblinded --> iffalse)
\iffalse
\author{
\IEEEauthorblockN{Anonymous Author(s)}
}
\else
\author{
\IEEEauthorblockN{Michael Baddeley\IEEEauthorrefmark{1},
Carlo Alberto Boano\IEEEauthorrefmark{2},
Antonio Escobar-Molero\IEEEauthorrefmark{3},
Ye Liu\IEEEauthorrefmark{4},
Xiaoyuan Ma\IEEEauthorrefmark{5}, \\
Usman Raza\IEEEauthorrefmark{1},
Kay R\"{o}mer\IEEEauthorrefmark{2},
Markus Schu{\ss}\IEEEauthorrefmark{2},
and
Aleksandar Stanoev\IEEEauthorrefmark{1}} \vspace{-2.50mm}\\
\IEEEauthorblockA{\IEEEauthorrefmark{1}Toshiba Europe Ltd., Bristol, United Kingdom -- \url{{michael.baddeley, aleksandar.stanoev, usman.raza}@toshiba-bril.com} }
\IEEEauthorblockA{\IEEEauthorrefmark{2}Institute of Technical Informatics, Graz University of Technology, Austria -- \url{{cboano,schuss,roemer}@tugraz.at} }
\IEEEauthorblockA{\IEEEauthorrefmark{3}RedNodeLabs UG, Munich, Germany -- \url{antonio@rednodelabs.com} }
\IEEEauthorblockA{\IEEEauthorrefmark{4}College of Engineering, Nanjing Agricultural University, China --  \url{yeliu@njau.edu.cn} }
\IEEEauthorblockA{\IEEEauthorrefmark{5}Shanghai Advanced Research Institute, Chinese Academy of Sciences, China --  \url{maxy@sari.ac.cn} } \vspace{-4.75mm}\\
}
\fi

% Make title
\maketitle

% Abstract (no more than 250 words)
\begin{abstract}
The popularity of concurrent transmissions (CT) has soared after recent studies have shown their feasibility on the four physical layers specified by \blefive, hence providing an alternative to the use of \ieee for the design of reliable and efficient low-power wireless protocols. % on multi-radio chips. 
However, to date, the extent to which physical layer properties affect the performance of CT has not yet been investigated in detail. 
This paper fills this gap and provides the first extensive study on the impact of the physical layer on CT-based \mbox{solutions using \ieee and \blefive}. 
\mbox{We first highlight through simulation} how the \mbox{impact of} errors induced by de-synchronization and beating on the performance of CT highly depends on the choice of the underlying physical layer. 
We then confirm these observations experimentally on real hardware through an analysis of the bit error distribution across received packets, unveiling possible techniques to effectively handle these errors. 
We further study the performance of CT-based flooding protocols in the presence of radio interference on a large-scale, and derive important insights on how the used physical layer affects their dependability. 
\end{abstract}

\iffalse
% Keywords
\begin{IEEEkeywords}
Beating effect, \blefive, concurrent transmissions, \dcubee, energy-efficiency, flooding, \ieee, nRF52840, performance, physical layer, radio interference, reliability.
\end{IEEEkeywords}
\fi

% Introduction
\section{Introduction} \label{sec:introduction}

% Introducing concurrent transmissions 
A recent breakthrough in the low-power wireless community has been the development of communication protocols based on Concurrent Transmissions (CT). 
CT-based solutions intentionally let multiple relaying nodes forward packets by simultaneously broadcasting them \mbox{on the same carrier frequency}.
Thanks to the capture effect~\cite{leentvaar76capture} and to non-destructive interference~\cite{liao2016revisiting}, nodes overhearing these concurrent transmissions have a high probability to receive at least one transmission correctly, which enables the creation of reliable and efficient cyber-physical systems and Internet of Things (IoT) applications~\cite{zimmerling20synchronous}.  
%CB: @MB: CPS is never used as abbreviation and removing the acronym saves us one line :-)
%MB: @CB: Super!

The key benefit of CT is the ability to exploit sender diversity to realize simple flooding and synchronization services across large-scale multi-hop wireless networks~\cite{ferrari2011efficient}, as well as improving wireless performance in single-hop systems.
%MB: I've added single-hop to empasise this is a general wireless approach and should not be pigeonholed as only for multi-hop mesh.
%UR: I removed single hop because there are not any well-known single-hop CT-based protocols.
%CB: we had a long discussion with Michael about this and all agreed on it. 
Relaying nodes in a mesh network utilizing CT-based protocols do not need to explicitly avoid collisions using conventional techniques such as carrier sensing, % and scheduled access to the medium,
%\ms{well technically its still scheduled, maybe even more so}
%MB: @MS: I get your point, but I think it's pretty aparent what we mean. Will see if the reviewer picks up on it. If they do it's not going to sink the paper.
and can avoid the overhead of routing and link-based communication~\cite{zimmerling20synchronous}. 

A large body of work has proposed CT-based data collection~\cite{ferrari2012low, suzuki13choco, istomin2016data} and dissemination~\cite{doddavenkatappa2013splash, du17pando} protocols that can achieve unprecedented gains in terms of reliability, end-to-end latency, and energy efficiency. 
These protocols can outperform existing solutions even in the presence of harsh radio interference~\cite{lim2017competition, istomin18crystal, ma20harmony, escobar2019competition}, as shown by four editions of the EWSN dependability competition~\cite{boano17competition}.

However, the vast majority of CT-based protocols have only been implemented and verified experimentally using off-the-shelf platforms based on 2.4\,GHz \ieee radios, e.g., the very popular but rather outdated TelosB mote~\cite{polastre05telos}.  
\mbox{These solutions employ} a physical layer (PHY) based on Orthogonal Quadrature Phase Shift Keying (OQPSK) and Direct Sequence Spread Spectrum (DSSS), as specified by the \ieee standard~\cite{802154_ieee}, where DSSS provides the coding robustness needed by CT to be sufficiently reliable~\cite{wilhelm14concurrent}. 

\boldpar{CT and the impact of different PHYs}
An experimental study by Al Nahas et al.~\cite{alnahas2019concurrentBLE5} has shown the feasibility of CT also when using Bluetooth Low Energy~(BLE).
Their preliminary results show that reliable and efficient \mbox{CT-based} flooding is possible on BLE-based mesh networks, but highlight that the performance \emph{largely depends on the employed PHY}, as confirmed by the measurements reported in Schaper's MSc thesis~\cite{schaper2019truth}. 
Indeed, the most recent version of Bluetooth Low Energy~(\blefive) supports four PHYs that largely differ in terms of data rate and robustness~\cite{spoerk19ble5phy}: 2M (2\,Mbps), which doubles the nominal throughput of the original 1M\,PHY (1\,Mbps), and two coded PHYs with coding rates of 1/2~and~1/8 respectively \mbox{(i.e., the 500K and 125K\,PHYs}). 
% These results suggest that coded PHYs may help in increasing the reliability of CT when using BLE. 

These preliminary observations are important in light of the increasing number of commodity IoT platforms that embed low-power radios supporting multiple wireless standards and PHYs, as they hint that developers need to carefully select the physical layer used for CT-based communication.  
Notable examples of such off-the-shelf platforms are the TI~CC2652R~\cite{cc2652_productsheet}
%\ms{would this paper still apply to the newer cc2652rb with the baw?} 
%MB: @MS: Removed as I think we talked about this and concluded no issue.
and the Nordic Semiconductors nRF52840~\cite{nrf52840_productsheet}, which support, among others, the 2.4\,GHz \ieee OQPSK-DSSS PHY as well as the four PHYs specified by \blefive on the same chip. 

However, to date, \emph{the extent to which physical layer properties affect CT-based solutions employing \ieee and \blefive has not yet been investigated in detail}. 
Firstly, there is no experimental study systematically analyzing how physical layer effects such as beating\footnote{Beating is a pulsating interference pattern between two or more signals at slightly different frequencies, as described in Sect.~\ref{sec:background}.}, induced by relative carrier frequency offset~\cite{liao2016revisiting}, and de-synchronization due to clock drift~\cite{escobar19imprel} affect the reliability of the received signal in the presence of multiple concurrent transmitters. 
Furthermore, there is no experimental work trying to verify whether the robustness of CT-based flooding protocols to radio interference holds true when using different PHYs (or studying whether the performance of CT in harsh RF environments differs depending on the used PHY). 

Shedding light on these aspects is important to (i)~provide a better understanding on the role of the physical layer on the reliability and efficiency of CT, as well as to (ii)~empower developers to use the physical layer as a means to fine-tune the performance of CT-based protocols at runtime.

% Figure from background
\begin{figure}[!t]
	\centering
	\vspace{-2.50mm}
	\includegraphics[width=0.80\columnwidth]{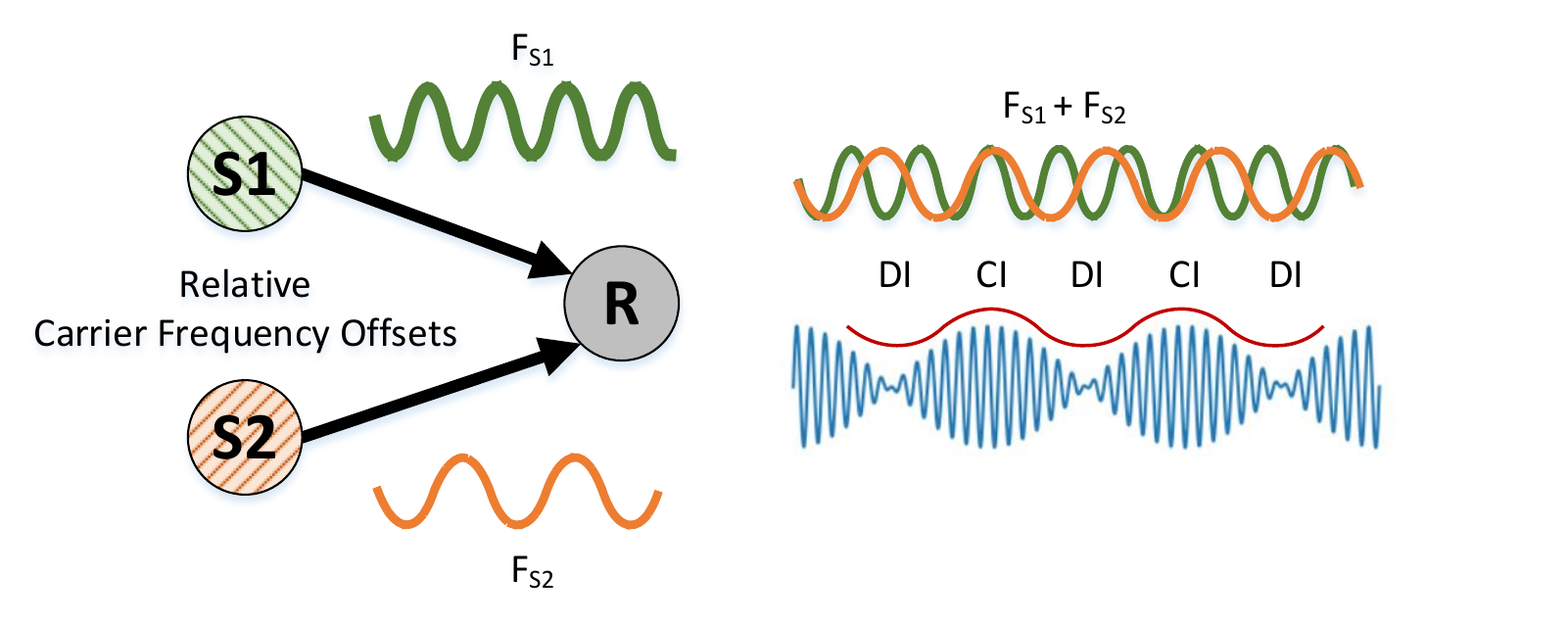}
	\vspace{-2.50mm}
	\caption{Beating due to relative oscillator frequency inaccuracies between devices. Signals combine to produce periods of \emph{constructive} and \emph{destructive} interference (CI~and~DI). }
	\vspace{-4.00mm}
	\label{fig:ct-beating-background}
\end{figure}

\boldpar{Our contributions}
This paper addresses this gap and provides the first in-depth experimental study on the impact of the PHY on CT-based solutions employing \ieee and \blefive. 

We first simulate the performance of CT for the different \blefive PHYs, highlighting the role of beating under different interference scenarios.
% the impact that the various techniques used in different PHYs have on communication performance under beating conditions.
We then set up an extensive experimental campaign to confirm these simulation results and to systematically study the performance of CT across all the \ieee and \blefive PHYs supported by the nRF52840 platform. 
%\ms{i think we omitted the prop mode, so no shockburst, not sure what gazelle and ant run on}
%MB: @MS: I've now made it specific to ieee and BLE5.
To this end, we use the \dcubee public testbed~\cite{schuss17competition, schuss18benchmark}, recently enhanced with $50$ nRF52840-DK devices, to observe both beating frequencies and de-synchronization effects on real hardware through an analysis of the error distribution across received packets. 
Our experiments demonstrate that the impact of errors induced by de-synchronization and beating on CT performance is highly dependent on the choice of the underlying PHY, on the relative carrier frequency offset between transmitting devices, and on the number of concurrent transmitters. 
Specifically, we observe that: (i)~high data rate PHYs experience wider beating and can mitigate its impact through repetition; (ii)~if the power delta between signals is insufficient, then the \blefive convolutional coding is no longer effective to sustain reliable CT; (iii)~the pattern mapper used in the 125K PHY allows it to effectively handle narrow beating.

We further use \dcubee to perform the first experimental study on the performance of different CT-based flooding protocols as a function of the underlying PHY in the presence of RF interference on a large scale. 
To this end, we make use of \dcubee's \jamlabng functionality~\cite{schuss19jamlabng} to generate artificial \wifi interference and stress-test the performance of CT-based protocols such as Glossy~\cite{ferrari2011efficient} and robust flooding (RoF)~\cite{lim2017competition} under \emph{no}, \emph{mild}, and \emph{strong} interference. 
Our results allow us to derive important insights on which PHYs are effective to help CT-based protocols in mitigating the impact of interference. 
Such insights include: (i)~the superiority of \ieee and \blefive500K\,PHY under strong interference, (ii)~the fact that the BLE 125K\,PHY should not be used in conjunction with long payload lengths under interference, as well as (iii)~the need to dynamically change PHY at runtime to provide the best trade-off between reliability, latency, and energy efficiency.

% Summary of contributions 
\fakepar{}
After providing some background knowledge on CT in Sect.~\ref{sec:background}, this paper makes the following specific contributions:
\begin{itemize}
    \item We simulate the performance of CT for all \blefive PHYs, highlighting the role of beating (Sect.~\ref{sec:simulation}).
    \item We are the first to experimentally observe beating frequencies and de-synchronization effects on real hardware and wireless channel for different PHYs through an analysis of the bit error distribution across received packets (Sect.~\ref{sec:beating}).
    \item We evaluate the performance of CT-based flooding protocols under radio interference, and provide insights on how the employed PHYs affect dependability (Sect.~\ref{sec:interference}). 
\end{itemize}
We then describe related work in Sect.~\ref{sec:related_work} to highlight how our insights align with existing literature, and conclude our paper in Sect.~\ref{sec:conclusions} along with a discussion on future work.

% Background section
\section{Concurrent Transmissions in Low-Power\\ Wireless Networks} \label{sec:background}

\begin{figure}[!t]
	\centering
	\vspace{-1.00mm}
	\includegraphics[width=0.70\columnwidth]{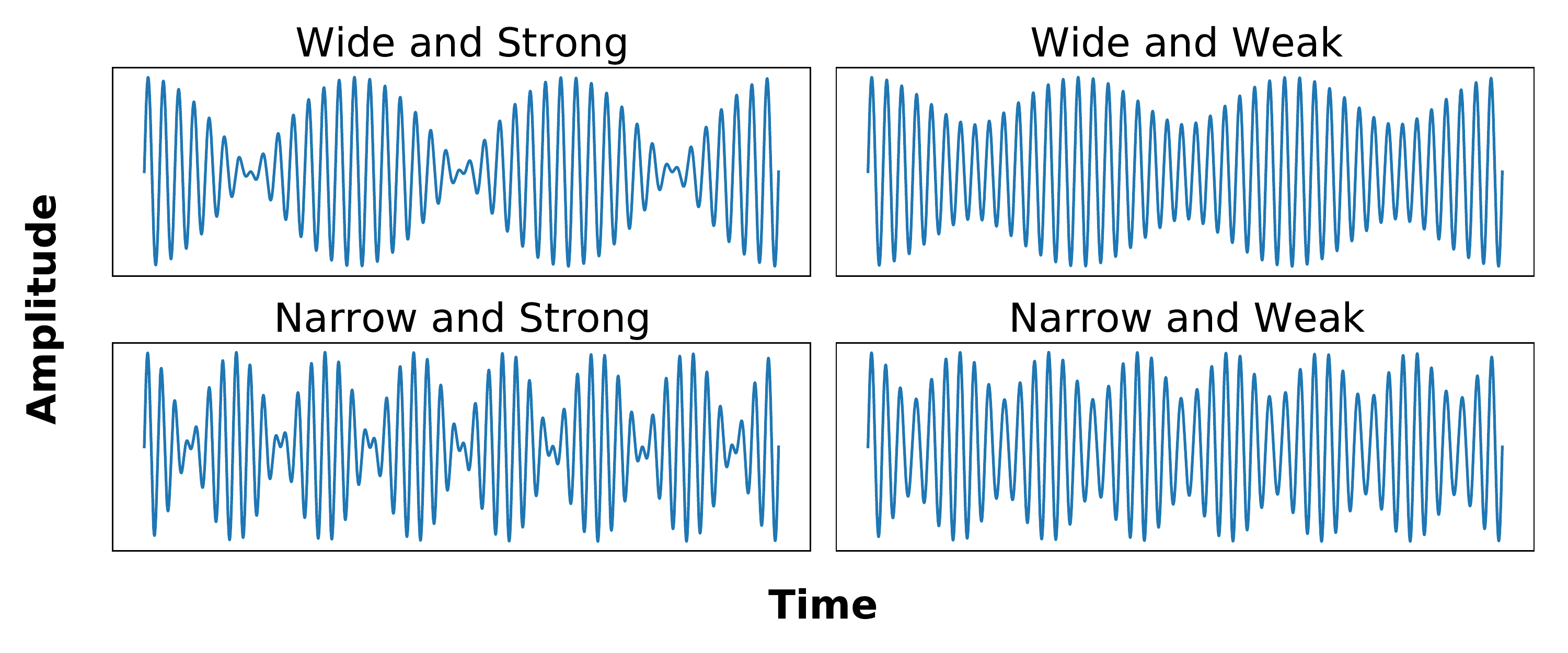}
	\vspace{-1.75mm}
	\caption{Beating manifests in many forms depending on the relative frequencies between devices and their channel gains. Strong\,beating\,attenuates\,a\,signal\,toward\,zero\,in\,a\,beating\,`valley'.}
	\label{fig:ct-beating-types}
	\vspace{-4.55mm}
\end{figure}

CT is the concept by which several nodes transmit the data they want to share at the same time. The physical layer of low-power IoT devices is typically based on different variations of binary frequency-shift keying (BFSK) modulation, as specified in both \blefive and \ieee\footnote{OQPSK with half-sine pulse shaping is equivalent to Minimum-Shift Keying (MSK) and can be demodulated as a frequency modulation~\cite{pasupathy1979minimum}.}. Early works such as Glossy~\cite{ferrari2011efficient} showed that, when using frequency-based modulations, if nodes are sufficiently synchronized, then transmissions of the \emph{same data} will align and the packet will be correctly received with cooperative gain. Meanwhile, later works have shown that transmissions of \emph{different data} greatly benefit from capture effect due to energy diversity between transmitters. CT therefore constitute a robust technique to deploy simple, diverse, and latency-optimal mesh networks. Nevertheless, recent literature has shown that CT introduce two types of errors that degrade the communication performance:
% The concurrent transmission of the same data leads to a beating radio signal, where the signal magnitude alternates between peaks and valleys instead of being uniform, as illustrated in Fig. X.
\begin{enumerate}[leftmargin=+5.75mm]
	\item \emph{Synchronization errors}. Concurrent transmitters are not perfectly synchronized, which introduces intersymbol interference when different bits overlap on the air. To minimize this effect, packet transmissions must be triggered within a time interval ideally lower than half the symbol period~\cite{escobar19imprel}.
	\item \emph{Beating Effect}. When CT overlap on the air, the resulting waveform has a beating amplitude due to alternating periods of constructive and destructive interference\,\,(beating). As shown in Fig.~\ref{fig:ct-beating-background}, with \emph{two} concurrent transmitters, the waveform's envelope has a sinusoidal shape, while featuring more complex forms when \emph{more than two} transmissions overlap~\cite{alnahas2020blueflood}. While potentially introducing a certain degree of energy gain during peaks, beating greatly increases the chances of bit errors during low-energy periods (valleys) and has generally a negative net effect. 
\end{enumerate}

\noindent Notably, beating will impact dense topologies when there is no dominant transmission and will consequently be affected by deep fading. 
On the other hand, its impact is reduced when different transmissions are received with enough energy diversity and the capture effect kicks in. 
In Fig.~\ref{fig:ct-beating-types} we categorize beating as \emph{wide and strong}, \emph{wide and weak}, \emph{narrow and strong}, or \emph{narrow and weak}. 
Beating will be randomly \emph{narrow} or \emph{wide} depending on the relative carrier frequency offset between transmitters, while it will manifest as \emph{strong} or \emph{weak} depending on the relative difference in received signal energy.

% Simulation
\section{Impact of Physical Layer on CT Performance: Analysis and Simulation} \label{sec:simulation}
While synchronization errors can be greatly reduced by properly designing the CT network protocol and its retransmission strategy, \textit{beating cannot be avoided}. Beating always appears when signals from non-coherent transmitters overlap in the air, due to their different carrier frequency offset (CFO). 
Moreover, the temporal period of the beating is different for each set of concurrent transmitters and randomly depends on their oscillator inaccuracies. The unpredictable temporal length of the beating largely affects the error rate, since the beating periods can be very narrow (and several peaks and valleys can appear within a packet transmission), or very wide (and a packet can be completely shadowed within a valley).

\begin{figure*}[!t]
	\vspace{-1.75mm}
	\begin{subfigure}[t]{1\columnwidth}
		\centering
		\includegraphics[width=.67\columnwidth]{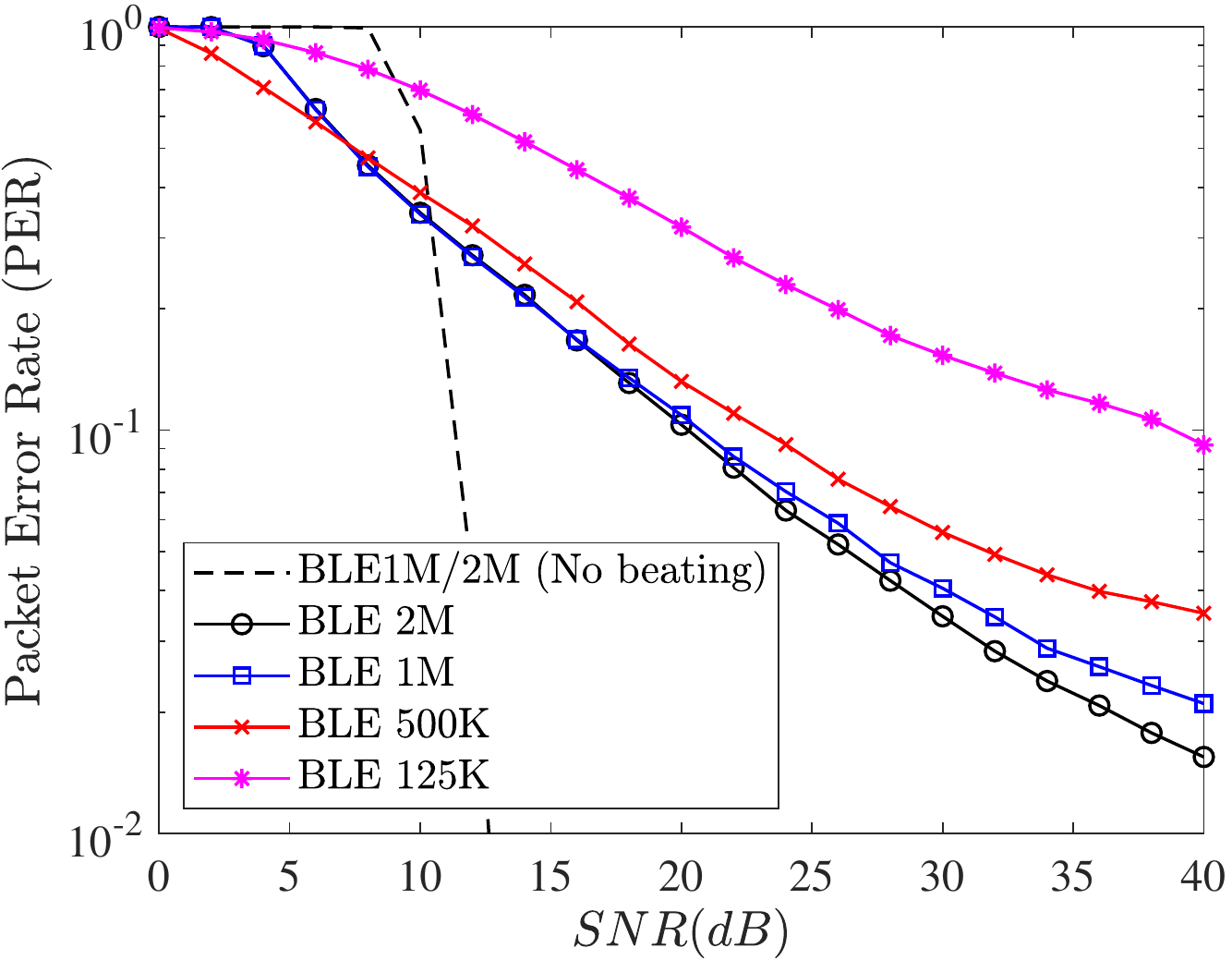}
		\vspace{-1.00mm}
		\caption{Wide and strong beating (2 CT, RFO=500Hz, $\Delta$P=0dB)}
		\label{fig:rednode_sim1}
	\end{subfigure}%
	\hfill
	\vspace{-1.75mm}
	\begin{subfigure}[t]{1\columnwidth}
		\centering
		\includegraphics[width=.67\columnwidth]{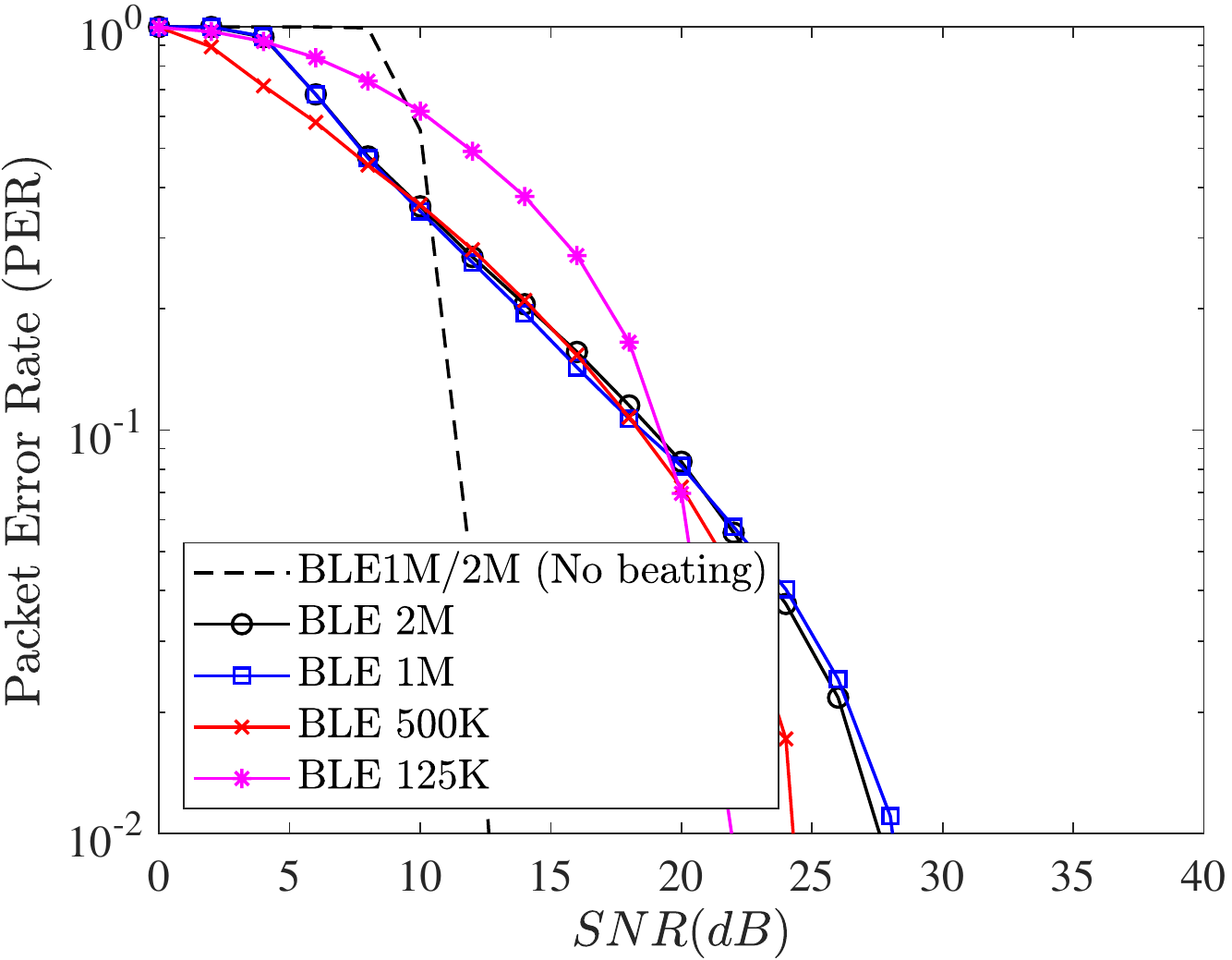}
		\vspace{-1.00mm}
		\caption{Wide and weak beating (2 CT, RFO=500Hz, $\Delta$P=1dB)}
		\label{fig:rednode_sim2}		
	\end{subfigure}%	
    \vskip\baselineskip
    \vspace{-2.00mm}
	\begin{subfigure}[t]{1\columnwidth}
		\centering
		\includegraphics[width=.67\columnwidth]{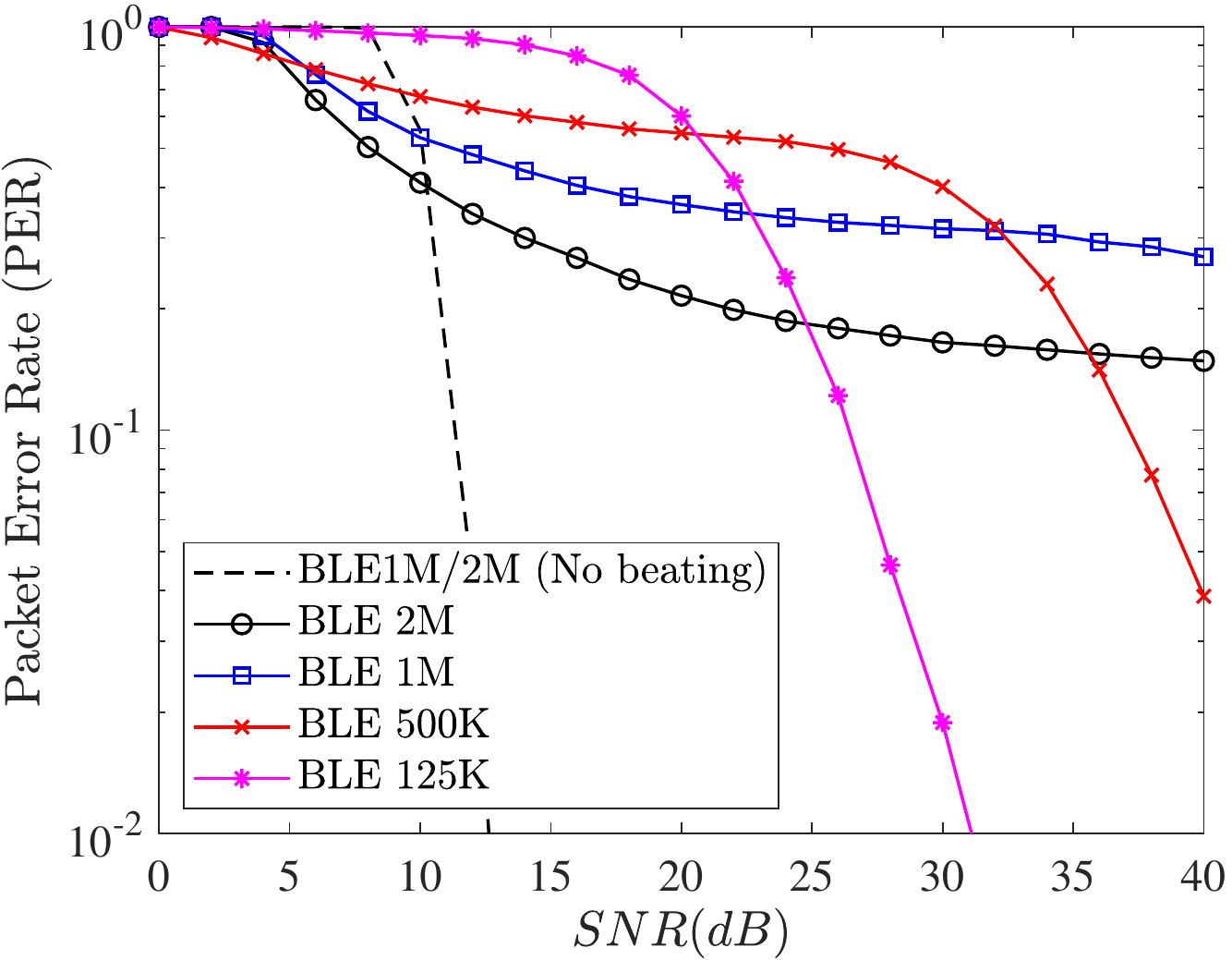}
		\vspace{-1.00mm}
		\caption{Narrow and strong beating (2 CT, RFO=10kHz, $\Delta$P=0dB)}
		\label{fig:rednode_sim3}		
	\end{subfigure}%
    \hfill
	\begin{subfigure}[t]{1\columnwidth}
		\centering
		\includegraphics[width=.67\columnwidth]{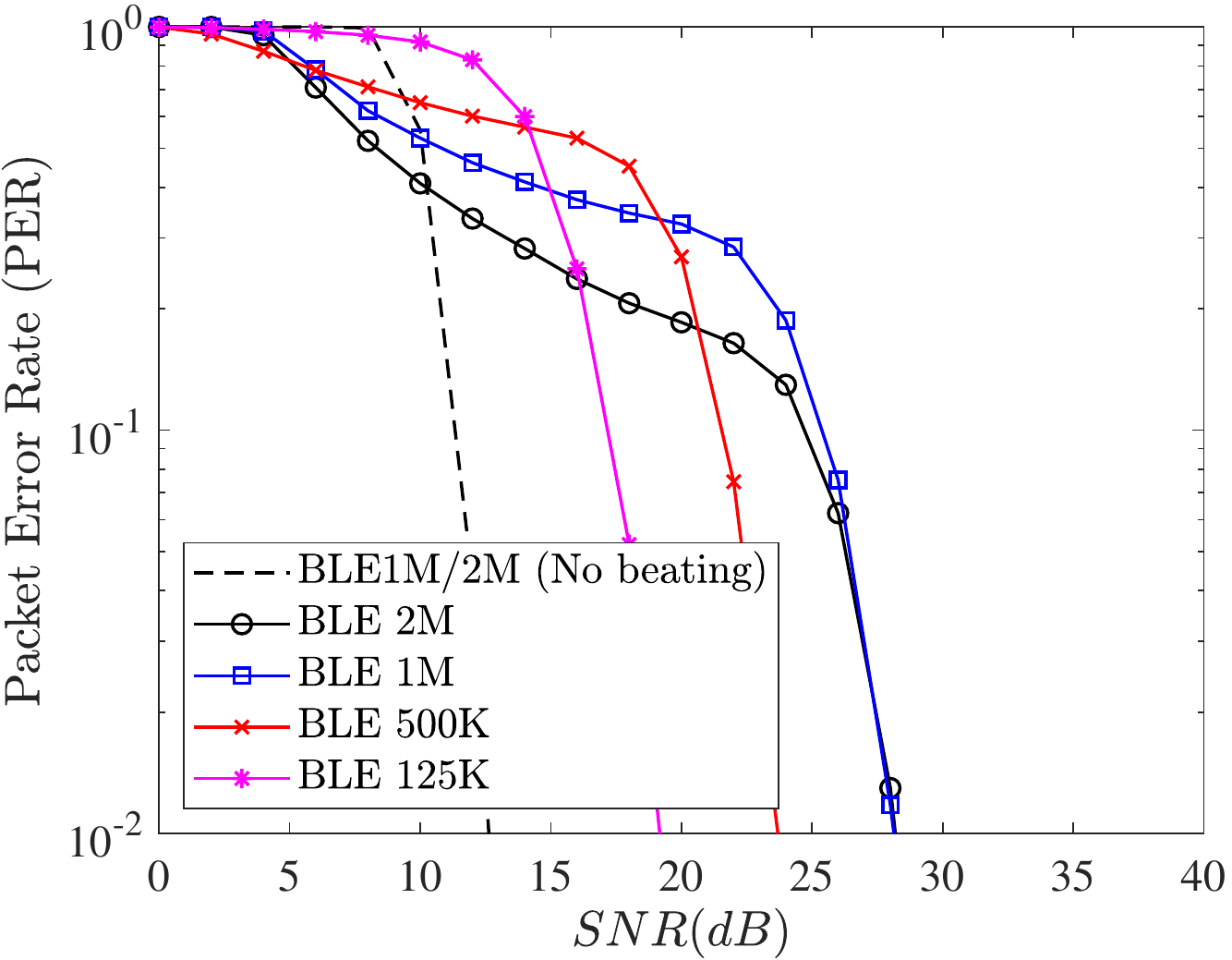}
		\vspace{-1.00mm}
		\caption{Narrow and weak beating (2 CT, RFO=10kHz, $\Delta$P=1dB)}
		\label{fig:rednode_sim4}		
	\end{subfigure}%
	\vspace{-1.25mm}
	\caption{Simulation results showing the PER when sending 30-byte packets with two concurrent transmitters and with a single transmitter (no beating). Differences in beating periods (wide or narrow) and power deltas have a significant impact on how beating affects performance, which explains why different node pairs (with different RFOs) may experience very different PERs.}
	\label{fig:rednode_sim}
	\vspace{-4.75mm}
\end{figure*}

%It is complex to analytically predict the exact impact that different techniques used in the PHYs have in the communication performance under beating conditions. To better analyze this, we simulate the different communication systems using MATLAB to obtain statistically meaningful average Packet Error Rate (PER) vs. Signal-to-Noise Ratio (SNR) for two CT recovered with a non-coherent BFSK receiver, as in \cite{escobar20phd}, assuming constant Additive White Gaussian Noise~(AWGN) and no synchronization errors. 
To better analyze the impact that different PHYs have on the performance of CT under beating, we simulate the different communication systems using MATLAB to obtain %statistically meaningful 
the average Packet Error Rate (PER) vs. Signal-to-Noise Ratio (SNR) for two CT recovered with a non-coherent BFSK receiver, as in~\cite{escobar20phd}. 
We assume constant additive white Gaussian noise and no synchronization errors. 
We repeat this for the different \blefive coded (500K and 125K) and uncoded (1M and 2M) PHYs, for different oscillator inaccuracies (which result in either wide or narrow beating) and power deltas. Both coded PHYs are based on the 1M PHY, adding a convolutional code of rate 1/2 and are received with a hard-decision Viterbi decoder~\cite{viterbi67code}. In addition, the 125K PHY adds a Manchester pattern mapper of four elements per coded bit\footnote{When using \blefive125K PHY's Manchester Pattern Mapper, a `0' is translated into `0011', whereas a `1' is translated to `1100'~\cite{ble5specs, ble5distance}.}.

We define the Relative Frequency Offset (RFO) -- which determines the beating frequency -- as the difference between the CFO of each individual transmitter, and the power delta $\Delta$P as the power ratio with which both CT are received. The SNR is defined relative to the strongest transmission (assuming $P_{R1}>P_{R2}$) and $N$ being the noise power: \vspace{-3.00mm} \\
\begin{equation}
RFO(Hz) = |CFO_1 - CFO_2| = 1 / T_{Beating},
\end{equation}
\vspace{-6.00mm}
\begin{equation}
\Delta P(dB) = 10\log_ {10}(P_{R1}/P_{R2}),
\end{equation}
\vspace{-6.00mm}
\begin{equation}
SNR(dB) = 10\log_ {10}(P_{R1}/N).
\end{equation}

\vspace{-1.00mm}

The BLE standard requires the CFO to be within $\pm$150\,kHz~\cite{ble5specs}, which results in RFOs lower than 300\,kHz. Therefore, the RFO is always lower than the (coded or uncoded) symbol frequency (i.e.,  2\,MHz in \blefive2M and 1\,MHz for the other three PHYs).
%(1\,MHz in the coded PHYs and \blefive1M, and 2\,MHz in \blefive2M).
%UR: To increase the clarity... too many Ms

\noindent The results of our simulation are presented in Fig.~\ref{fig:rednode_sim}. Based on these results, we derive the following observations: 

\begin{enumerate}[leftmargin=+5.75mm]
\item \textbf{Impact of beating.} We first compare the results obtained with two concurrent transmitters (2~CT) with those obtained with a single transmitter (no beating). With low-noise (SNR\,$>$\,15\,dB), beating negatively affects packet reception and increases the PER. Only when operating in high-noise conditions (SNR\,$<$\,10\,dB), 2~CT experience a PER lower than that of a single transmitter, due to the positive net effect of constructive interference intervals. CT are hence an optimal mechanism in harsh environments with high noise, in which packet loss is high. Otherwise, the effect of destructive interference dominates and the PER increases.

\item \textbf{Wide ($\boldsymbol{T_{Beating}$\,$>$\,$T_{Packet}}$) and strong ($\boldsymbol{\Delta P$\,$\approx$\,0\,$dB}$) beating, Fig.~\ref{fig:rednode_sim1}}. In this case, $T_{Packet}$, which denotes the over-the-air time of a packet\footnote{$T_{Packet}$ is computed as $T_{Packet} = B \cdot (1 / DR)$, with $DR$ being the data rate of the chosen PHY and B being the length of the packet in bits.}, is the key factor dictating the PER. Indeed, the probability that the transmission spans a destructive interference interval is lower as the time the packet spends on the air decreases. Hence, when subjected to a same fixed $T_{Beating}$, uncoded PHYs (\blefive1M and 2M) perform better than coded ones (125K and 500K), since the former benefit from faster transmissions. With wide energy valleys, convolutional codes are ineffective: as a result, \blefive125K is the worst performing PHY, since it features the longest packet durations.

\item \textbf{Narrow\,($\boldsymbol{T_{Beating}$\,$<$\,$T_{Packet}}$)\,and\,strong\,($\boldsymbol{\Delta P$\,$\approx$\,0\,$dB}$) beating, Fig.~\ref{fig:rednode_sim3}}. In this configuration, the packet transmission always spans one or more destructive valleys. The \blefive2M PHY benefits from having the shortest packet duration, outperforming the 1M PHY. The convolutional encoder used in the \blefive500K and 125K PHYs is ineffective against beating, since it is optimal for discretely distributed one bit errors, but not to correct the burst errors that typically occur with beating. Nevertheless, the \blefive125K PHY features a good performance in narrow beating conditions when noise is very low (SNR\,$>$\,20\,dB), experiencing a waterfall-like PER decrease, whereas \blefive500K does not experience such a decrease until SNR\,$>$\,30\,dB, performing worse than uncoded PHYs in the mid-to-low noise range.

\item \textbf{Very low noise (SNR\,$>$\,25\,dB) and strong beating ($\boldsymbol{\Delta P$\,$\approx$\,0\,$dB}$), Fig.~\ref{fig:rednode_sim1} and Fig.~\ref{fig:rednode_sim3}}. Under these conditions, the most promising PHYs are \blefive2M, when fast data rate is needed, and \blefive125K when long-range is required and a limited data rate is sufficient. With the \blefive2M PHY, in a flooding-based mesh network using CT, the optimal strategy is using several (fast) retransmissions to increase the chances of successful packet receptions and compensate the lower PER compared to classical (without beating) routing schemes. In dense networks, in which multiple (more than two) retransmitters with different RFOs are simultaneously in the range of the receivers, narrow beating conditions are more frequent, and therefore one can exploit the performance boost of the \blefive125K PHY, which potentially requires no retransmissions (in contrast to the use of \blefive2M PHY). However, if packets are too long, \blefive125K may enter the wide-beating region, in which it performs poorly, and may suffer from synchronization errors. Therefore, with CT, \blefive125K should not be used to transmit packets with a large payload.

\item \textbf{Very high noise (SNR\,$<$\,5\,dB), Fig.~\ref{fig:rednode_sim1} and Fig.~\ref{fig:rednode_sim3}}. In such configuration, the \blefive500K PHY performs well and represents the best choice to survive extremely noisy environments. Nonetheless, we believe that as the number of transmitters increases the properties of the system tend to be dominated by the internal interference caused by beating, decreasing the influence of external noise. %While not covered in these results, this should factor into any future work.
% AE: this is a concept we want to explore potentially in an extension, maybe we can remove it by now. 
%Therefore, dense networks using CT can be modelled equivalently to operating in the high-SNR region of the 2 CT model, while low-SNR operation appears only in sparse networks or extremely jammed scenarios.

\item \boldpar{Weak beating ($\boldsymbol{\Delta P}$\,$>$\,0\,dB), Fig.~\ref{fig:rednode_sim2} and~\ref{fig:rednode_sim4}} In real deployments, CT are normally received with dissimilar power levels. Under weak beating, the signal does not completely fade during the valleys, which greatly decreases the impact of beating on the PER. %, as can be seen in the simulations already with only 1\,dB difference. 
For greater dissimilarities ($\Delta P$\,$>$\,3\,dB), the capture effect kicks in, and the PER becomes very similar to the no-beating scenario. Under weak beating, the 1M PHY may also be a suitable alternative, especially in sparse networks where beating is wider. This is because it has a comparable performance to the 2M PHY, while being more tolerant to synchronization errors (it requires a synchronization within 0.5\,$\mu$s instead of 0.25\,$\mu$s), and having a better receiver sensitivity.
\end{enumerate}

In real-world networks, all four scenarios depicted in Fig.~\ref{fig:rednode_sim} may simultaneously appear in different sections of a multi-hop network, depending on the practically unpredictable CFOs and power level relationships between the concurrent transmitters. Even for a given link, conditions may change over time, since temperature alters the CFO and surrounding interference or multipath propagation cause dynamic fluctuations in the power deltas. It is hence desirable that an optimal PHY tailored to handle CT features robust behavior in all four scenarios, since controlling operating conditions within a wireless network is challenging. In particular, within dense mesh networks with potentially many more than two CT at every hop, narrow and strength-varying beating is expected to dominate, and links can be modelled as behaving like the narrow beating figures for 2~CT (Fig. \ref{fig:rednode_sim3} and Fig. \ref{fig:rednode_sim4}) working in the high-SNR region.

\boldpar{Observations for beating mitigation} 
Based on these simulation results, we infer that beating can be mitigated by boosting energy diversity with proper node placement and techniques to dynamically control the transmission power. Nonetheless, this would affect the main advantages of using CT-based protocols: scalability and simplicity. When using lower data rates, which ultimately results in narrower beating periods relative to the packet period, it is more likely that packets will not completely fade: here, coding and Forward Error Correction (FEC) techniques can be exploited to introduce sufficient diversity to recover the errors introduced during the valleys by using the information received during the peaks. This is the case for IEEE 802.15.4, which uses Direct-Sequence Spread Spectrum (DSSS), and for the BLE coded PHYs (500K and 125K), which feature FEC. However, the convolutional encoder used in BLE coded PHYs is not effective in low-noise conditions, and the main gain in this region comes from the addition of the Manchester pattern mapper in \blefive125K. An exception is the case of networks operating in very noisy environments, in which the coded PHYs (especially \blefive500K) may be able to trigger successful packet receptions where classical (routing-based) schemes and coded PHYs may fail. Contrarily, higher data rates experience relatively wider beating periods, since the packet duration is shorter. In this scenario, whole packets may be blocked by wide valleys, which completely precludes any correction attempt. Similarly, the packet may randomly experience a wide peak, and be properly received. With high data rates, as in BLE uncoded PHYs (1M and 2M), the strategy should hence be different. Instead of trying to correct the errors, it would be more effective to repeat the packet several times, practically trying to randomly trigger a transmission spanning an energy peak.

\fakepar{}
Next, we confirm our simulation results on real hardware using \emph{over-the-air} experiments.  
As we expect physical layer properties to cause a fluctuating error probability across received packets, we directly observe beating by mapping the distribution of bit errors across a packet. 
%Sect.~\ref{sec:beating} presents the results of our experimental campaign. To the best of our knowledge, we are the first to demonstrate how physical layer effects such as RFO-induced beating affect the received signal using \emph{over-the-air} experiments. 

% Results from Toshiba
\section{CT Performance over different PHYs: Experimental Evaluation of Bit Error Distribution} \label{sec:beating}
Previous CT literature has attributed gains seen at the receiver to so-called Constructive Interference (CI)~\cite{ferrari2011efficient}. More recent works~\cite{liao2016revisiting, escobar19imprel}, in addition to this paper's analysis in Sect.~\ref{sec:simulation}, have proposed that, contrary to this assertion, instances of few concurrent transmitters will result in beating (observed as periodic peaks and valleys across a waveform) due to innate CFO inaccuracies between devices (defined as RFO in the previous section). Rather than a CI gain, beating causes periods of \emph{both} constructive \emph{and} destructive interference across the packet, leading to errors during beating valleys, while beating peaks will benefit from a receiver gain. 
This has recently been demonstrated experimentally by observing the raw in-phase and quadrature~(IQ) samples observed when connecting a small number of CT devices to a Software Defined Radio (SDR) using coaxial cables~\cite{alnahas2020blueflood}. Although these efforts help to better explain some of the processes underpinning CT-based communication, it is hard to directly witness and evaluate how the occurrence of fundamental physical layer properties affect CT performance.
%\ms{should we hint that newer radios like the nrf52833 could do this with 5.1 direction finding?}
%MB: @MS: Good point, but I don't think it's necessary here. Let's make a note of it and leave that to an expansion.
To date, there has been no \emph{over-the-air} testbed experiment able to demonstrate how PHY effects, such as RFO-induced beating and de-synchronization due to clock drift, directly affect the signal observed by a receiving node.

To address this gap, we present a set of experiments that evaluate physical layer effects on a 1-hop network of nodes communicating wirelessly by means of CT. 
Given that PHY properties cause the error probability to flux across the received packet (as shown in Sect.~\ref{sec:simulation}), we \emph{observe beating by mapping the distribution of bit errors across a packet} when considering a large transmission sample. 
Specifically, we study the CT performance across the multiple available PHYs supported by the nRF52840 devices in \dcubee. 
We observe both beating frequencies and de-synchronization effects through analysis of the received error distribution, and demonstrate that their impact on CT performance is highly dependent on the choice of the underlying PHY, on the RFO between transmitting devices, and on the number of concurrent transmitters. All experiments in this section use the \emph{Atomic-SDN} CT stack\footnote{Atomic-SDN employs a back-to-back CT approach similar to the Robust Flooding (RoF) protocol evaluated in Sect.~\ref{sec:interference}.}~\cite{baddeley2019atomic, baddeley2020thesis} developed for the EWSN 2019 Dependability Competition~\cite{baddeley2019competition}.

\begin{figure}[!t]
	\centering
	\vspace{-1.75mm}
	\includegraphics[width=0.63\columnwidth]{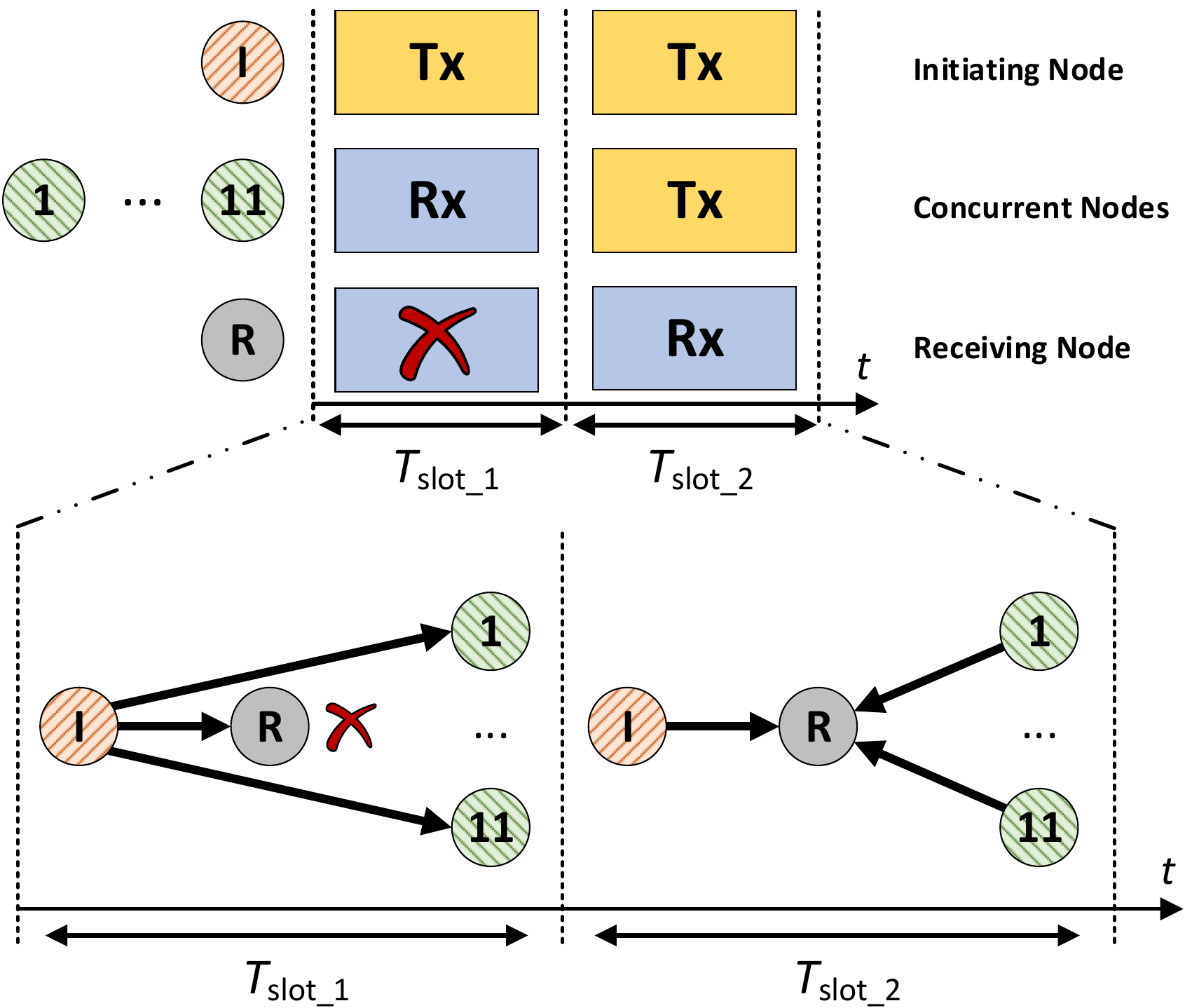}
	\vspace{-2.00mm}
	\caption{Experimental setup for examining 1-hop CT.}
%	 Selected concurrent nodes from $1-9$ synchronize their clocks in $T_{slot\_1}$, while the receiving node $R$ ignores this transmission. Both the initiating node $I$ and concurrent nodes then synchronously transmit to $R$ in $T_{slot\_2}$.
	\label{fig:beating_experimental_setup}
	\vspace{-4.00mm}	
\end{figure}

\begin{figure*}[ht]
	\vspace{-1.50mm}
	\centering
	\begin{subfigure}[t]{0.19\textwidth}
		\centering
		\includegraphics[width=1\columnwidth]{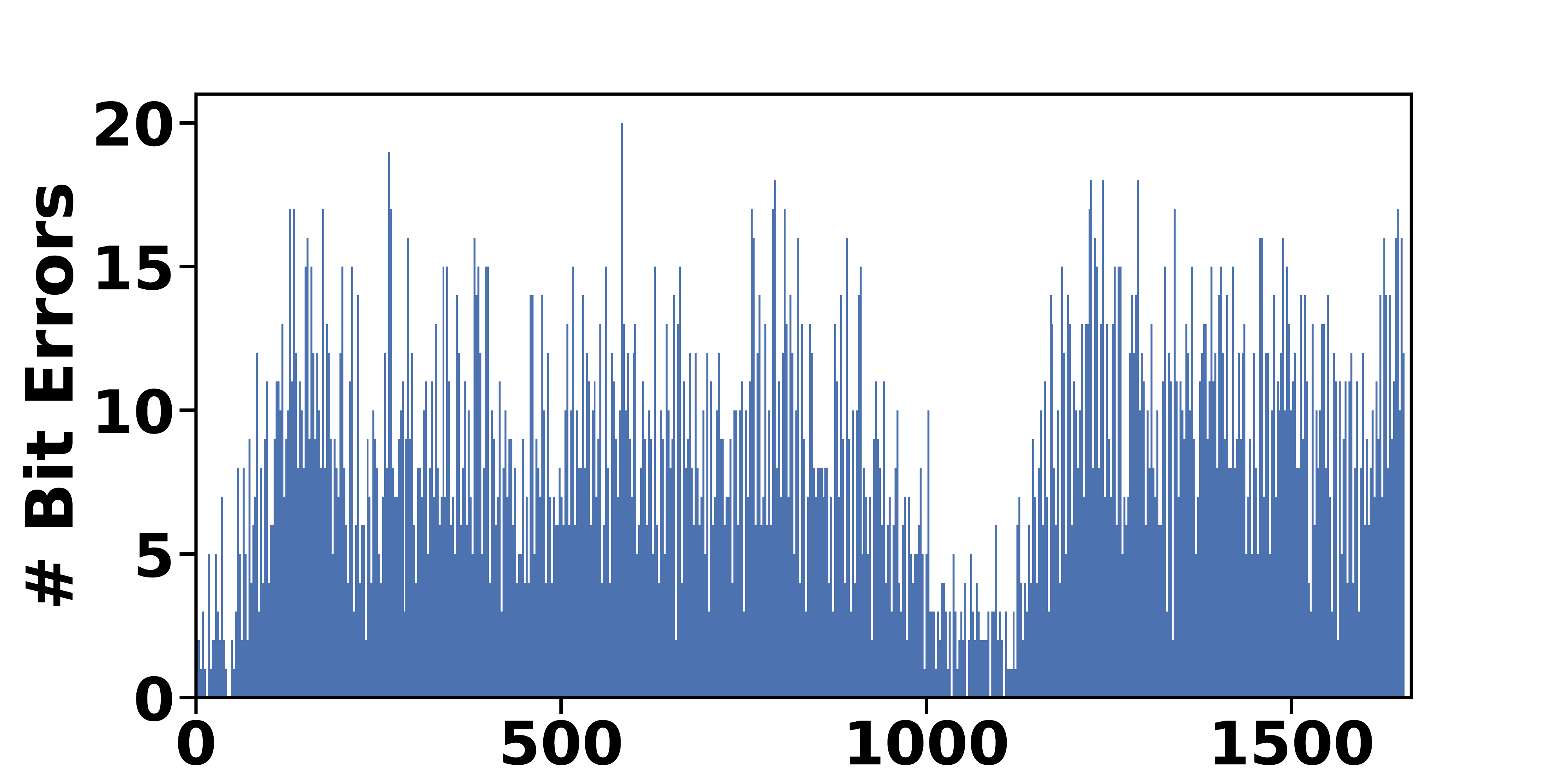}
		\vspace{-5.75mm}
		\caption{\blefive2M}
		\label{fig:tosh_phycomp_2m}
	\end{subfigure}%
	\begin{subfigure}[t]{0.19\textwidth}
		\centering
		\includegraphics[width=1\columnwidth]{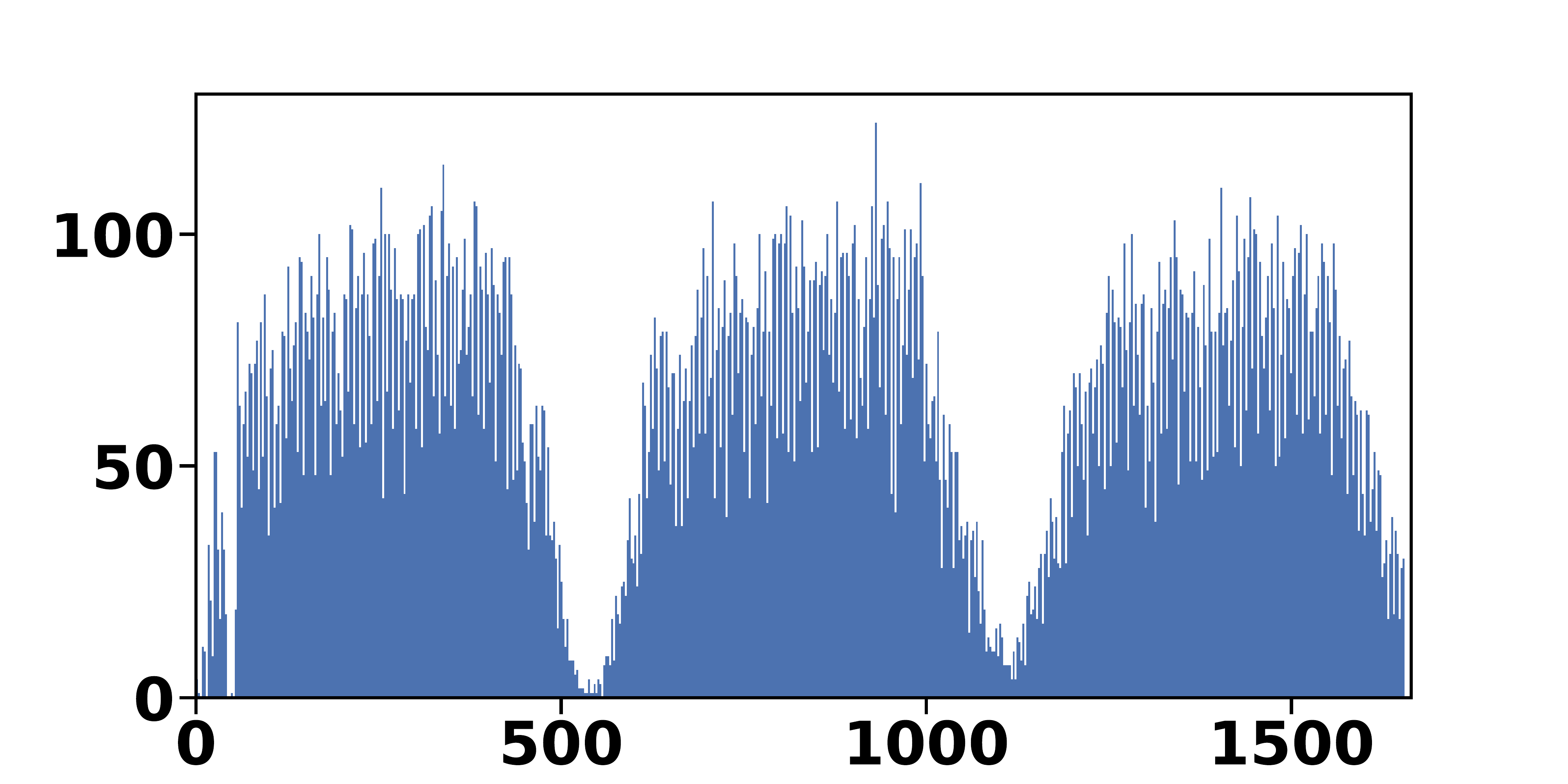}
		\vspace{-5.75mm}
		\caption{\blefive1M}
		\label{fig:tosh_phycomp_1m}
	\end{subfigure}
	\begin{subfigure}[t]{0.19\textwidth}
		\centering
		\includegraphics[width=1\columnwidth]{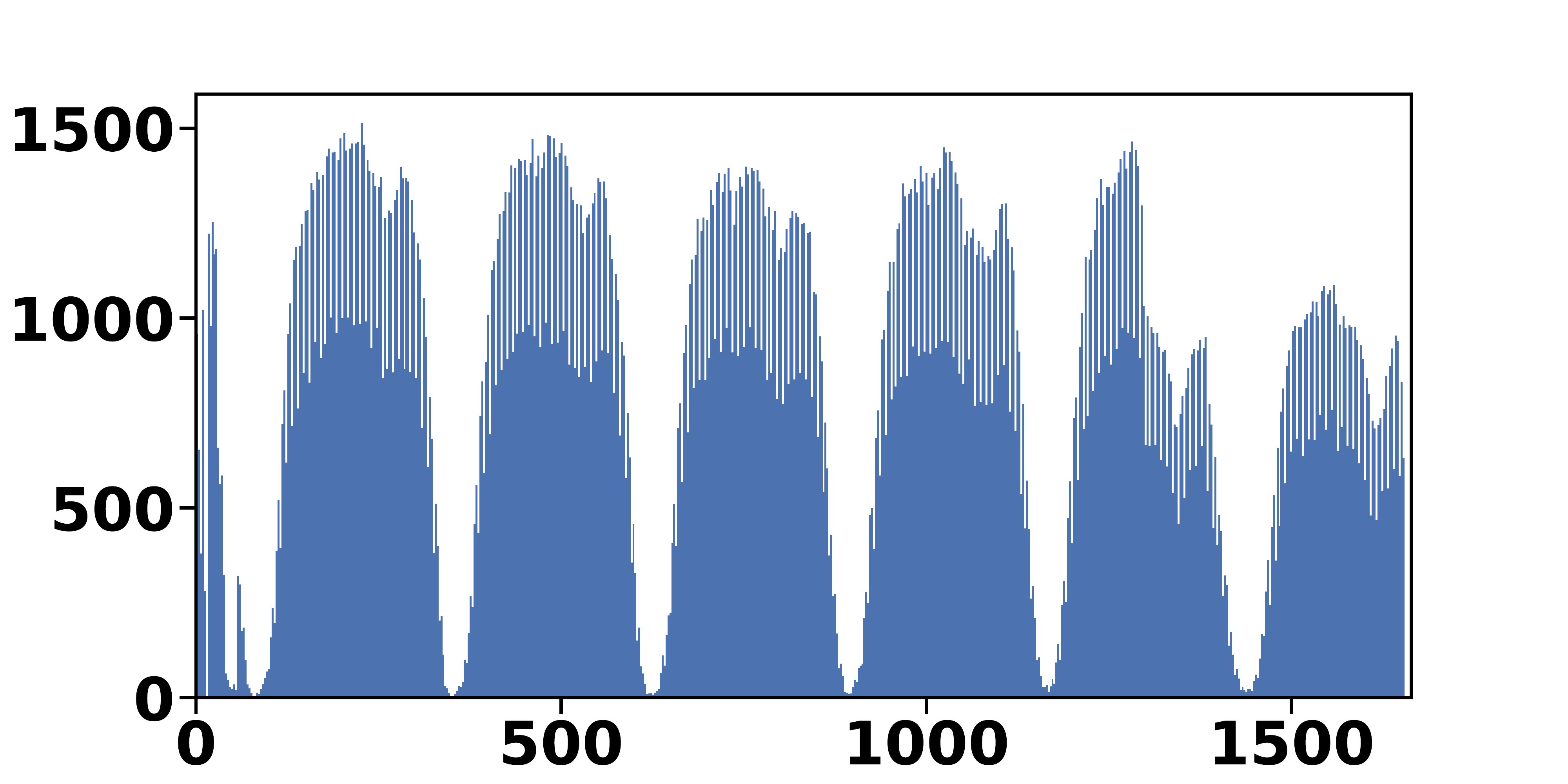}
		\vspace{-5.75mm}
		\caption{\blefive500K}
		\label{fig:tosh_phycomp_500k}
	\end{subfigure}
	\begin{subfigure}[t]{0.19\textwidth}
		\centering
		\includegraphics[width=1\columnwidth]{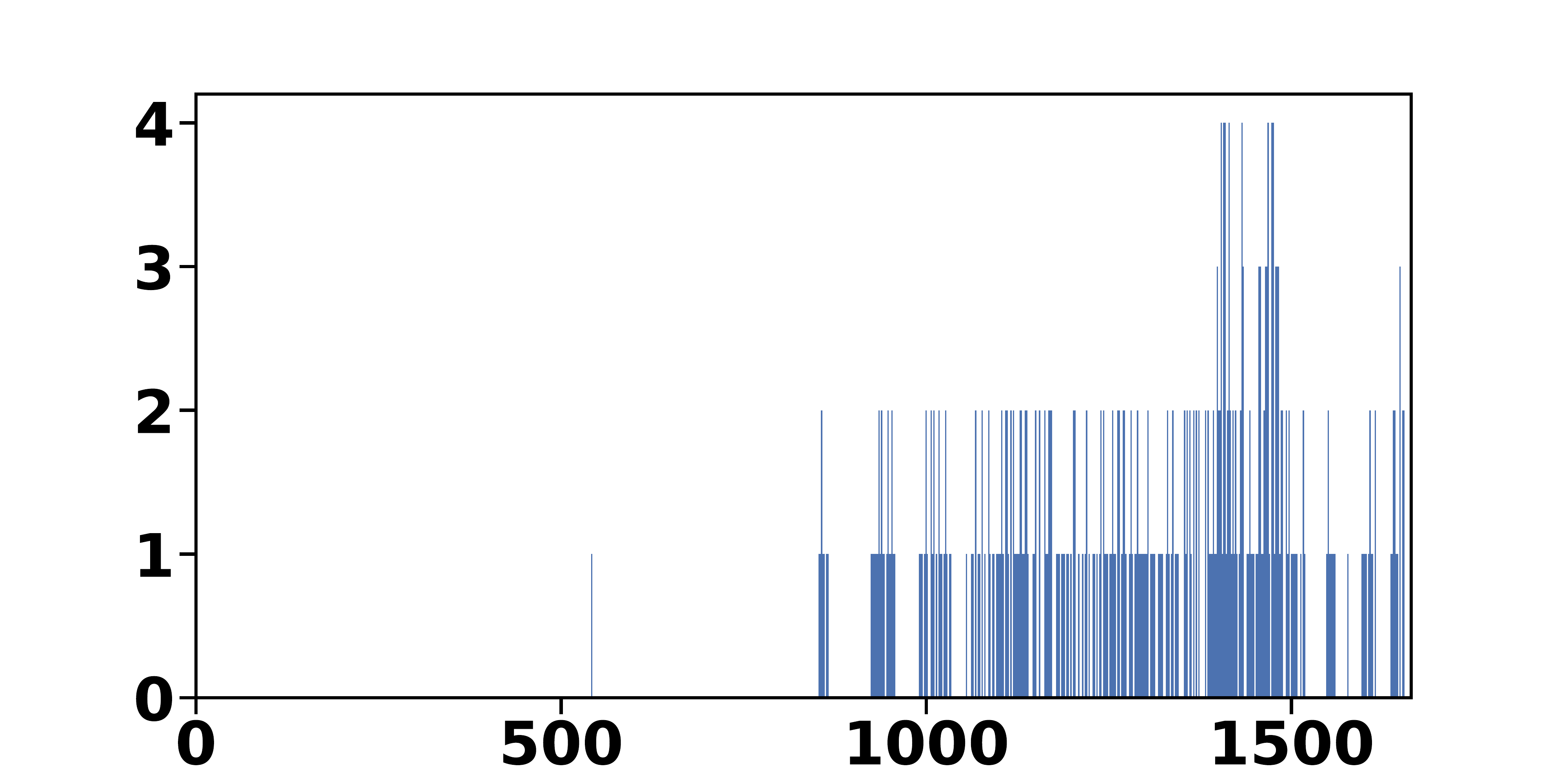}
		\vspace{-5.75mm}
		\caption{\blefive125K}
		\label{fig:tosh_phycomp_125k}
	\end{subfigure}
	\begin{subfigure}[t]{0.19\textwidth}
		\centering
		\includegraphics[width=1\columnwidth]{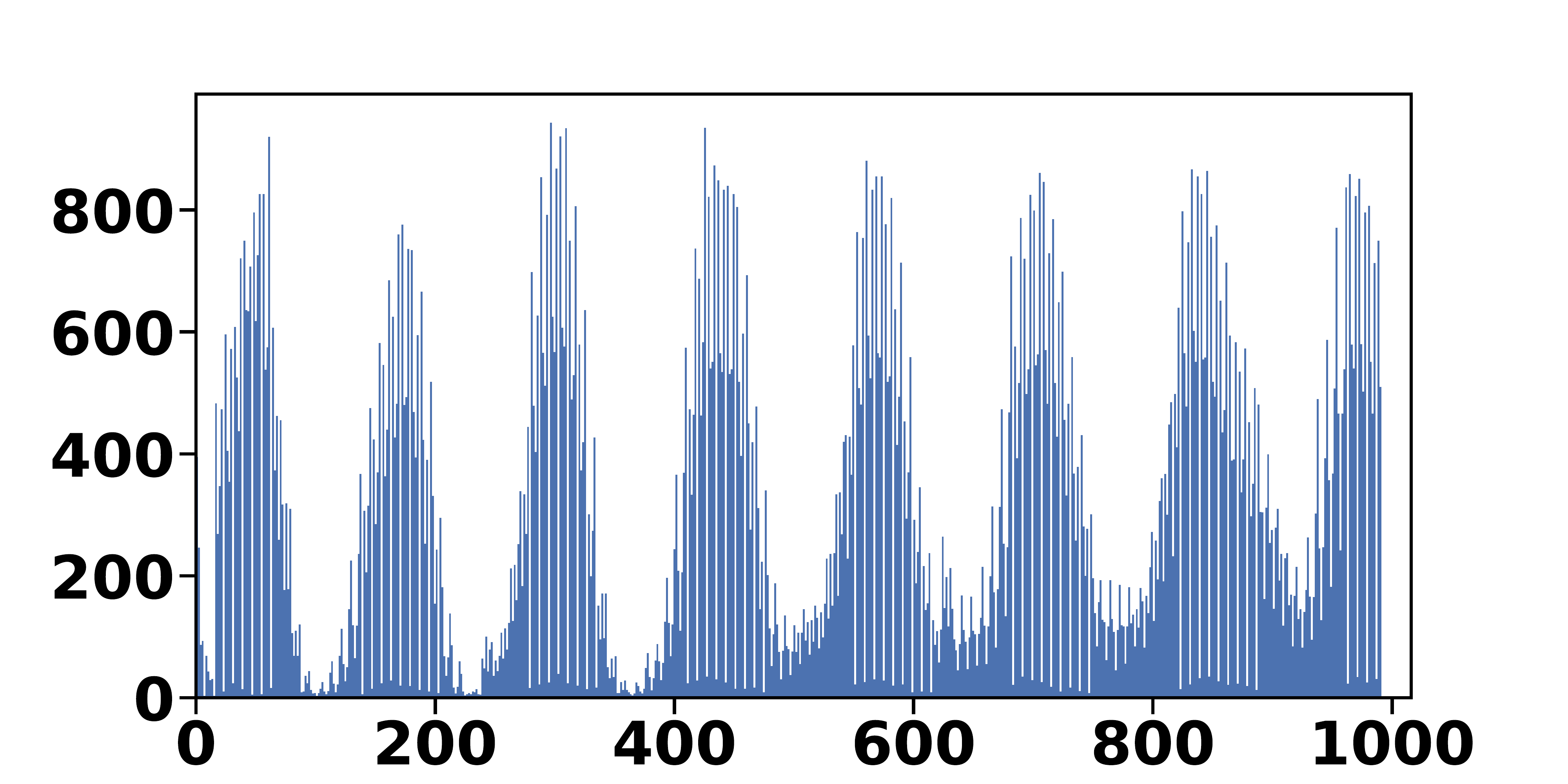}
		\vspace{-5.75mm}
		\caption{\ieee}
		\label{fig:tosh_phycomp_802154}
	\end{subfigure}
	\vspace{-2.25mm}
	\caption{Bit error distribution showing significant beating for a \emph{single} CT pair across all physical layers.}
	\label{fig:tosh_phycomp}
	\vspace{-1.00mm}
\end{figure*}

\begin{figure*}[ht]
	\centering
	\begin{subfigure}[t]{0.19\textwidth}
		\centering
		\includegraphics[width=1\columnwidth]{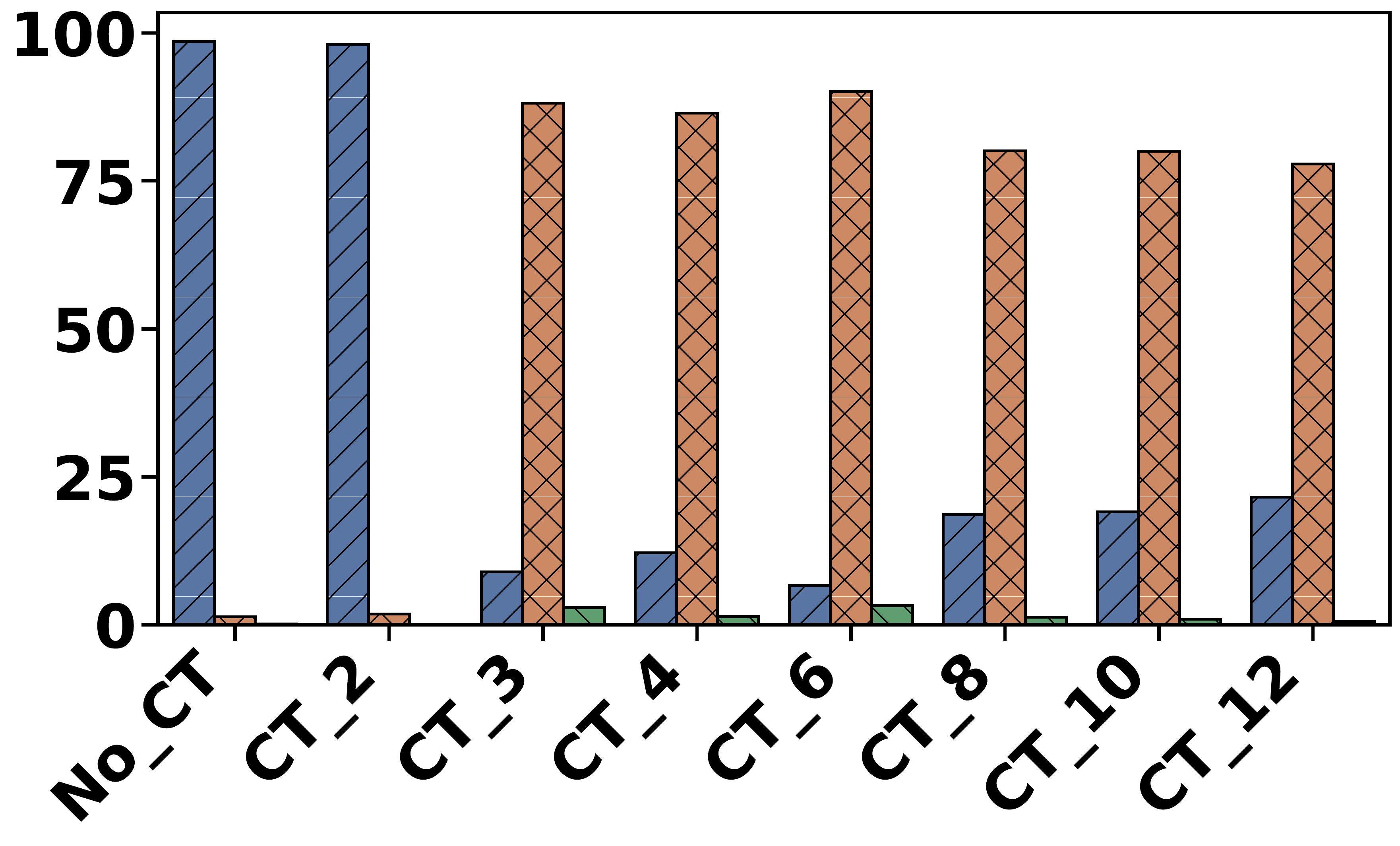}
		\vspace{-5.75mm}
		\caption{\blefive2M}
	\end{subfigure}%
	\begin{subfigure}[t]{0.19\textwidth}
		\centering
		\includegraphics[width=1\columnwidth]{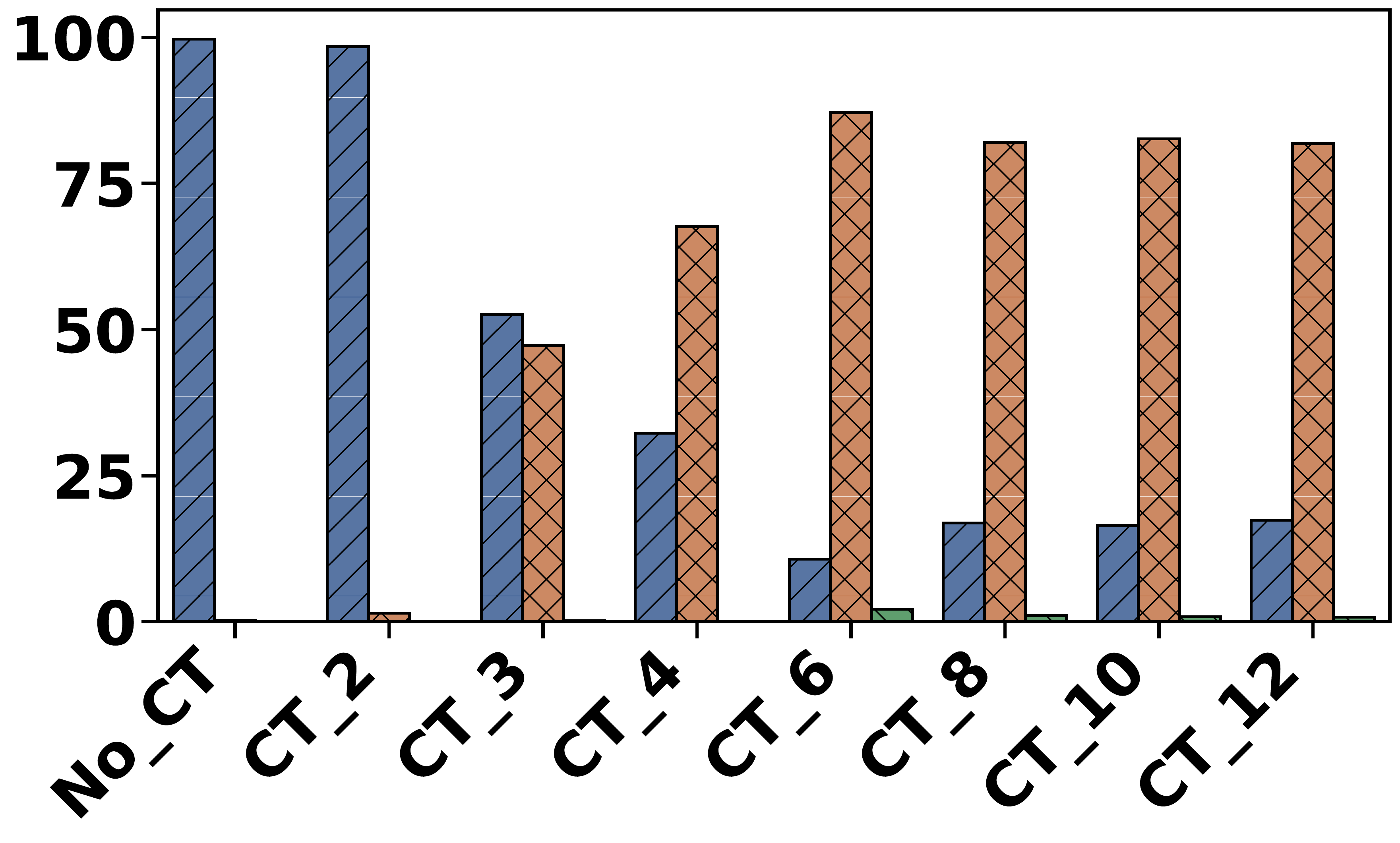}
		\vspace{-5.75mm}
		\caption{\blefive1M}
	\end{subfigure}
	\begin{subfigure}[t]{0.19\textwidth}
		\centering
		\includegraphics[width=1\columnwidth]{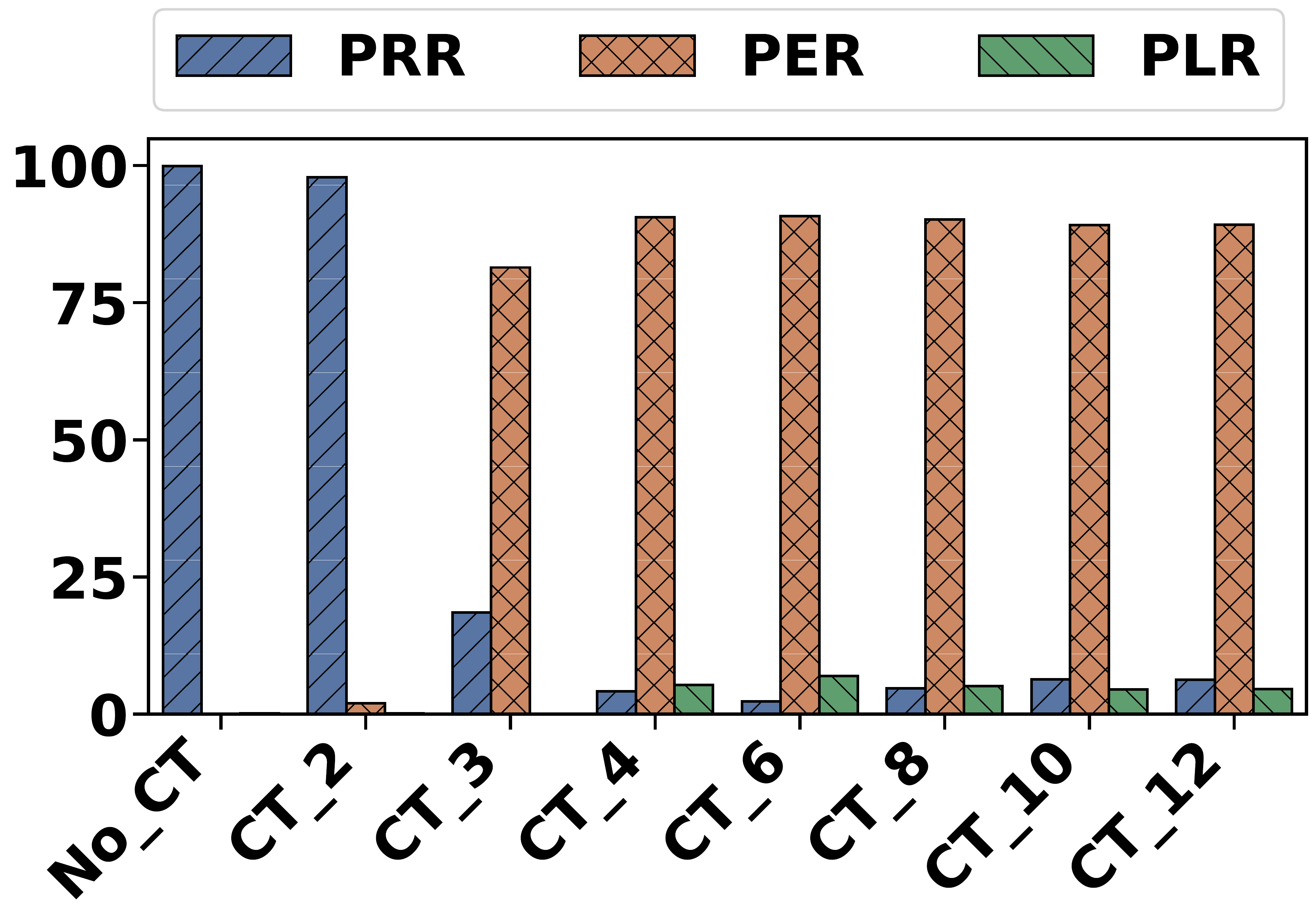}
		\vspace{-5.75mm}
		\caption{\blefive500K}
	\end{subfigure}
	\begin{subfigure}[t]{0.19\textwidth}
		\centering
		\includegraphics[width=1\columnwidth]{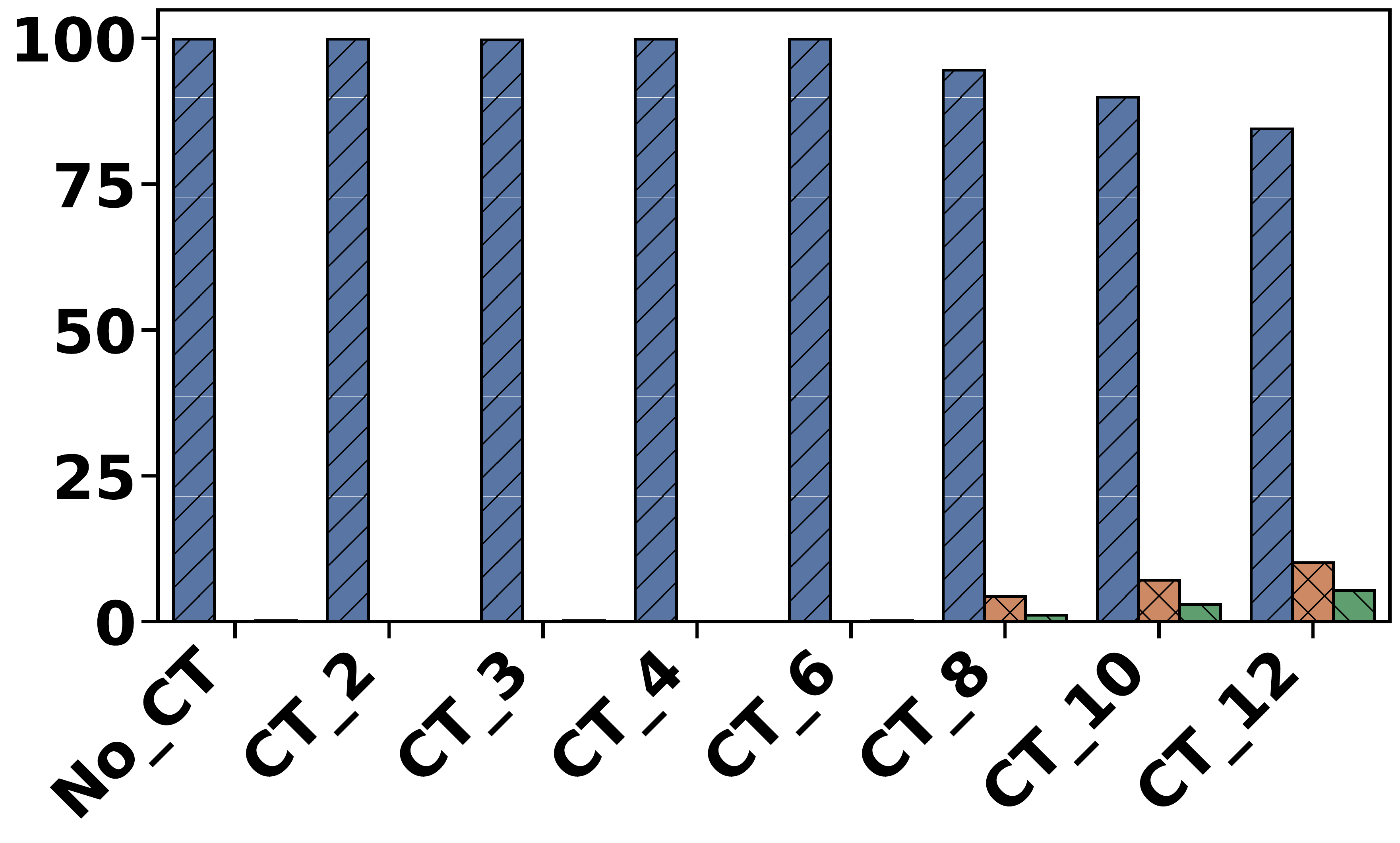}
		\vspace{-5.75mm}
		\caption{\blefive125K}
	\end{subfigure}
	\begin{subfigure}[t]{0.19\textwidth}
		\centering
		\includegraphics[width=1\columnwidth]{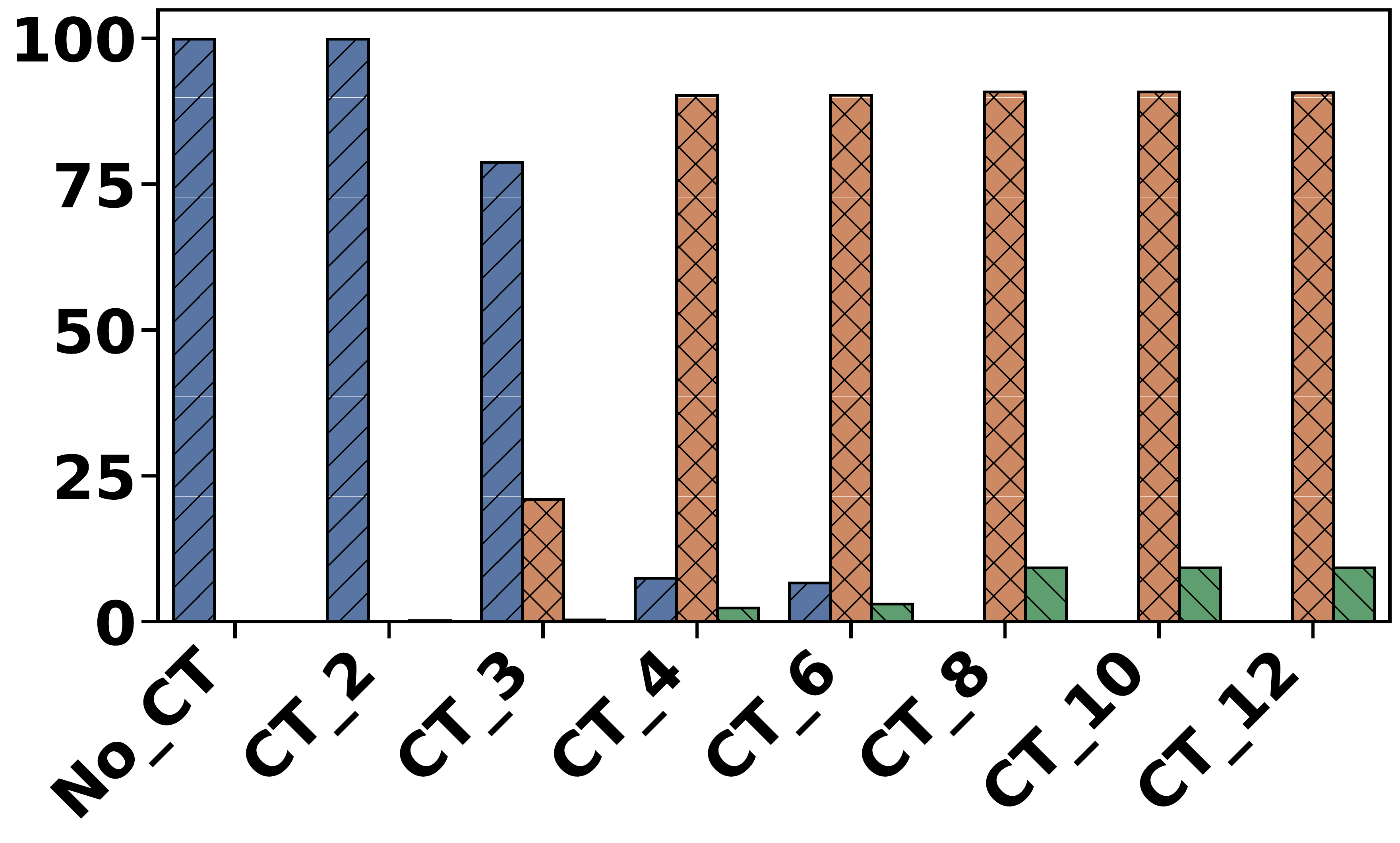}
		\vspace{-5.75mm}
		\caption{\ieee}
	\end{subfigure}
	\vspace{-2.25mm}
	\caption{PRR, PER, and PLR as CT density increases from a single transmitter (no CT) to CT12.}
	\label{fig:tosh_results_ct_comp}
	\vspace{-2.75mm}	
\end{figure*}

%---------------------------------------------------------%

% Experimental Design
%---------------------------------------------------------%
\subsection{Experimental Setup}
The \dcubee testbed was configured to provide a single-hop scenario for up to $12$ concurrently-transmitting nodes and a single fixed receiving node. Fig.~\ref{fig:beating_experimental_setup} demonstrates this setup and shows how the network is able to synchronize all transmitting nodes whilst limiting packet receptions at the receiver to only those that are a sum of signals from multiple concurrent transmitters. Node $R$ ignores the first transmission from the CT initiator $I$ in $T_{slot\_1}$, while allowing concurrent nodes to receive and synchronize to $I$. In $T_{slot\_2}$ all concurrent nodes synchronously transmit, including the initiator, and are observed at $R$. Node $I$ was configured to periodically generate and transmit a pseudo-random payload every 250\,ms, which was logged before each transmission in $T_{slot\_1}$, alongside an 8-byte CT header. 
When using \ieee, this was a 119B payload (due to the 127B maximum transmission unit limitation), while a payload of 200B was used for the \blefive PHYs\footnote{Although 200B is not the full maximum transmission unit of \blefive, it allows sufficient time to capture wider beating effects while reasonably limiting transmission time when using the \blefive125K PHY.}. At the receiving node $R$, the byte arrays of all correctly received % ($RX_{OK}$) 
and incorrectly received %($RX_{CRC}$) 
packets were logged in $T_{slot\_2}$. A direct comparison of transmitted and received byte arrays subsequently allows observation of the following four metrics.

\begin{enumerate}[leftmargin=+5.75mm]
	\item\emph{Bit error distribution} -- mapping the distribution of errors across a packet exposes observable periods of gain and interference over time. 
	\item\emph{Packet Reception Ratio (PRR)} -- indicating the ability of the physical layer to recover from CT interference. 
	\item\emph{Packet Error Ratio (PER)} -- indicating CT interference with a packet after the correct reception of the preamble.
	\item\emph{Packet Loss Ratio (PLR)} -- allowing the observation (by omission) of packets for which there was an energy minimum during a preamble's reception, resulting in the radio discarding the packet.
\end{enumerate}

\noindent This setup was run across the \ieee and all the \blefive PHYs for 2, 3, 4, 6, 8, 10, and 12 concurrently transmitters at -8\,dBm\footnote{A transmission power of -8\,dBm allows to capture sufficient bit errors within a reasonably short time window, minimizing experimentation time.}, where we define CT density as CT2, CT3,~...~CT12 respectively. Each experiment was run for $\approx$18K transmitted packets, representing over 100 total hours of experimental data. 

%---------------------------------------------------------%
% Experimental Results
%---------------------------------------------------------%

\subsection{Results}
\label{subsec:beating_results}

\begin{figure}[!t]
	\centering
	\vspace{-1.00mm}
	\includegraphics[width=0.65\columnwidth]{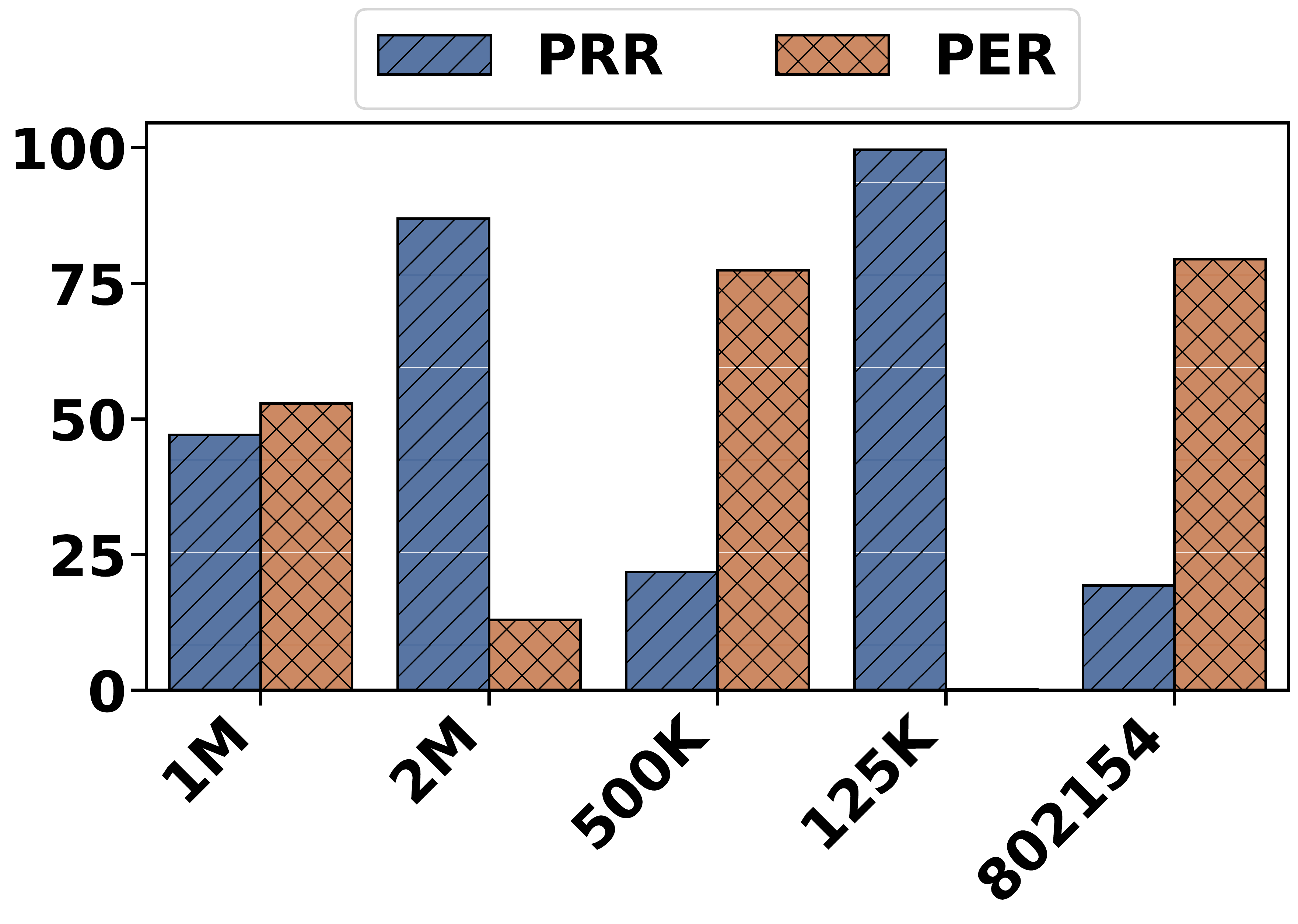}
	\vspace{-2.25mm}
	\caption{PRR and PER for the subplots shown in Fig. \ref{fig:tosh_phycomp}, where the CT2 pair experiences \emph{significant} beating over a 200B packet.}
	\label{fig:tosh_phycomp_ratio}
	\vspace{-0.2cm}
\end{figure}

%---------------------------------------------------------%
\boldpar{Beating effect on different PHYs} 
Fig.~\ref{fig:tosh_phycomp} shows the bit error distribution for a single CT2 pair across \ieee and all supported \blefive PHYs on the nRF52840. This \emph{specific} pair experiences significant beating, \emph{with valleys observed as bit error peaks} and clearly visible across all PHYs (with exception of the coded 125K). The sinusoidal waveform generated by this bit error distribution displays a constant beating period relative to the data rate for each PHY. These results are expanded further in Fig.~\ref{fig:tosh_phycomp_ratio}, which shows how the PRR and PER are closely linked to the way in which beating manifests across the various PHYs, and supports the findings previously shown through simulation in Fig.~\ref{fig:rednode_sim3} (\emph{narrow and strong} beating), where it is likely that this pair experiences \emph{very low noise} (SNR\,$>$\,25\,dB). As discussed in Sect.~\ref{sec:simulation}, the higher data rate PHYs (\blefive2M and 1M) experience fewer beating valleys. While these uncoded PHYs are unable to recover errors if they fall within a valley, the repetition commonly employed in CT protocols (\texttt{TX\_N\,=\,4} in these experiments) means it is likely that a retransmission will successfully fall between valleys and allow a successful reception of the preamble. This can be observed in the higher error rate for the 1M\,PHY, which experiences additional beating valleys as opposed to 2M. 
Furthermore, while it would be natural to assume that the redundancy employed in coded PHYs helps them to better recover from beating errors, our results show that the \blefive convolutional coding is unable to cope with significant beating (as seen from the high PER in the \blefive500K results shown in Fig.~\ref{fig:tosh_phycomp_ratio}). The same applies to the DSSS employed in \ieee, although it helps in recovering errors (despite the higher number of beating valleys caused by the lower data rate). On the other hand, the addition of the Manchester pattern mapper in the \blefive125K\,PHY provides sufficient gain to survive beating. %, and we start to witness de-synchronization errors due to very long transmission times.

%---------------------------------------------------------%
%MB: Removing this for space reasons. It's interesting but probably not necessary for this paper.
%\boldpar{RFO estimation} 
%Through examination of the bit distance between error peaks, it is possible to estimate the RFO of CT2 device pairs using Eq.~\ref{eq:tosh_rfo_calculation}, where $P_{dist}$ is the peak distance in bits, and $T_{bit}$ is the on-air time for a single bit for a given PHY. For example, an error peak distance of $\approx$ 500 bits at 1\,Mbps in Fig.~\ref{fig:tosh_phycomp_500k} gives a RFO of around 16\,KHz
%
%\begin{equation}
%RFO(Hz) = 1/(P_{dist} * T_{bit})
%\label{eq:tosh_rfo_calculation}
%\end{equation}

\begin{figure}[!t]
	\vspace{-1.50mm}
	\centering
	\begin{subfigure}[t]{0.3\columnwidth}
		\centering
		\includegraphics[width=1\columnwidth]{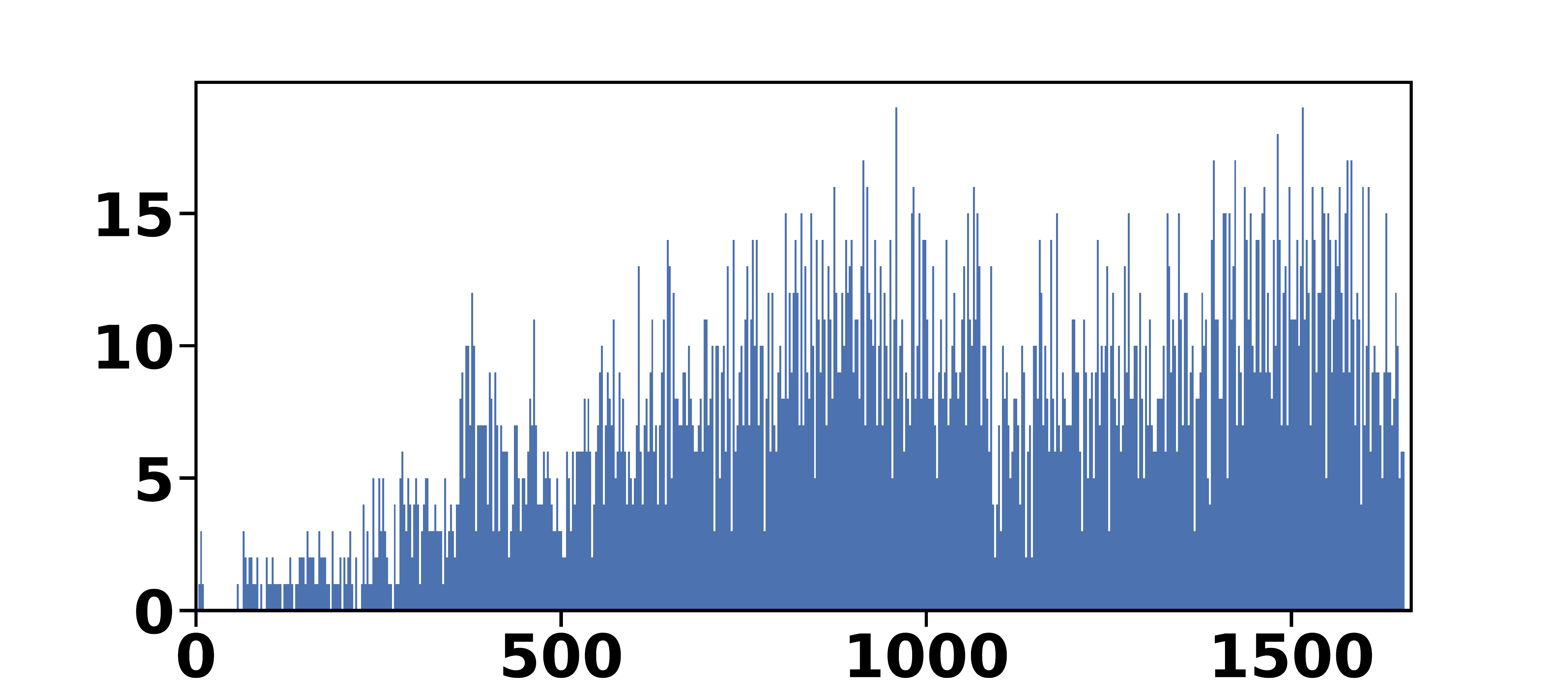}
		\vspace{-5.50mm}
		\caption{Pair 1}
		\label{fig:tosh_layoutcomp_1}
	\end{subfigure}%
	\begin{subfigure}[t]{0.3\columnwidth}
		\centering
		\includegraphics[width=1\columnwidth]{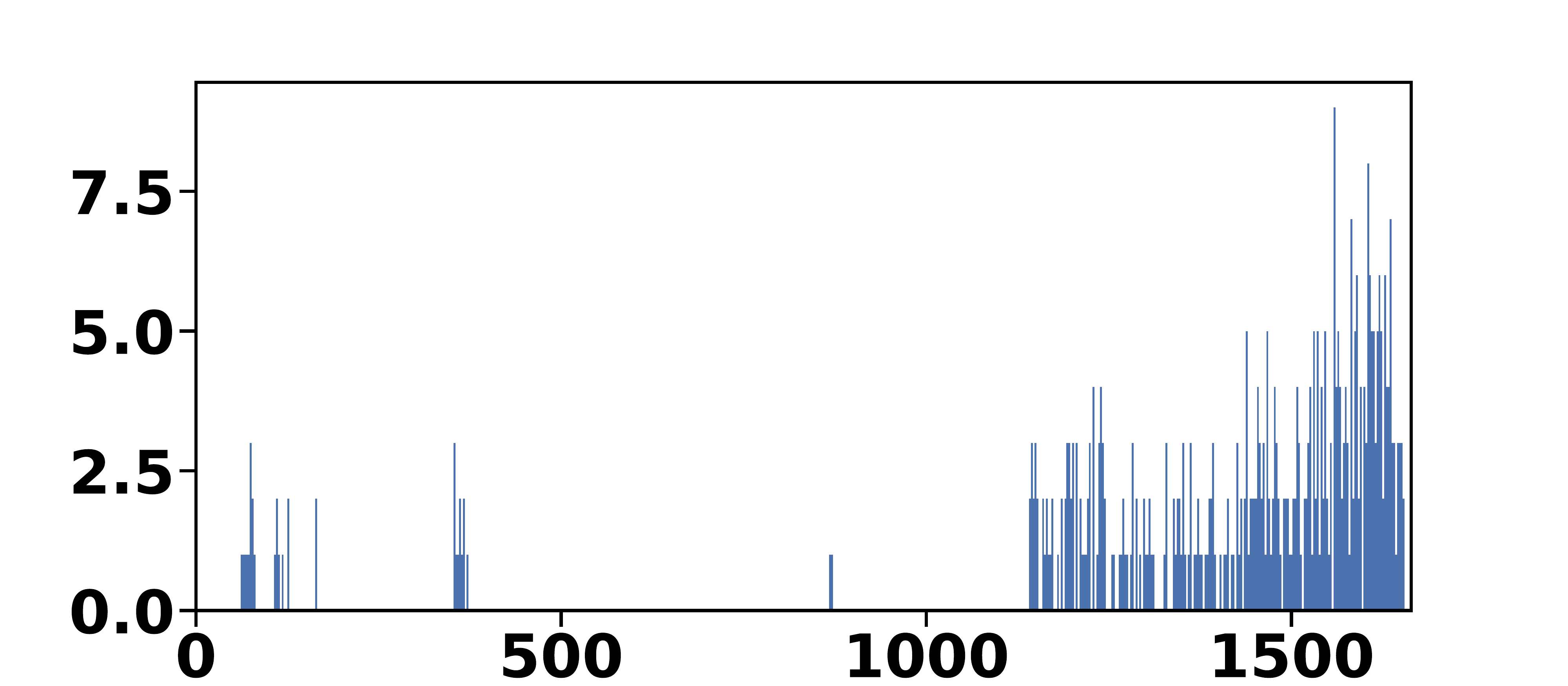}
		\vspace{-5.50mm}
		\caption{Pair 2}
		\label{fig:tosh_layoutcomp_2}
	\end{subfigure}
	\begin{subfigure}[t]{0.3\columnwidth}
		\centering
		\includegraphics[width=1\columnwidth]{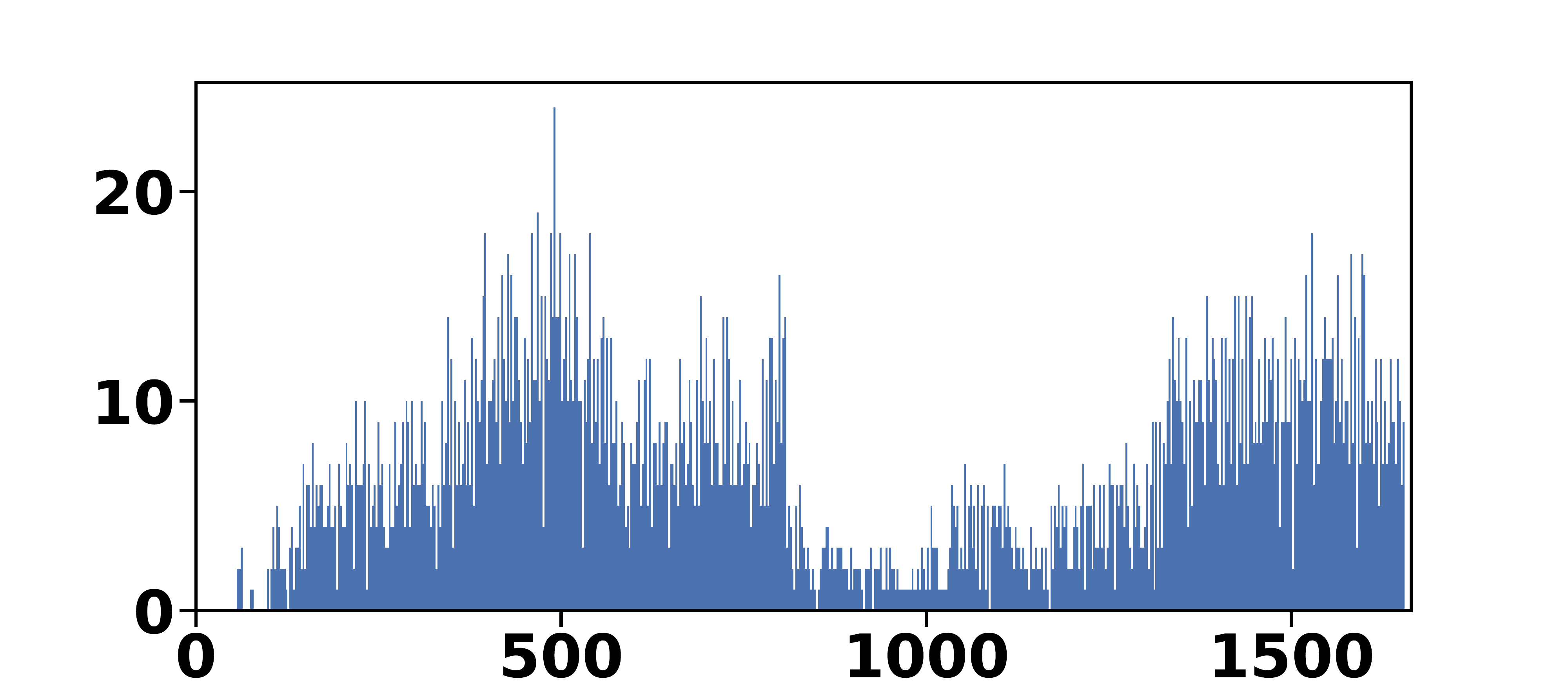}
		\vspace{-5.50mm}
		\caption{Pair 3}
		\label{fig:tosh_layoutcomp_3}
	\end{subfigure}
	\begin{subfigure}[t]{0.3\columnwidth}
		\centering
		\includegraphics[width=1\columnwidth]{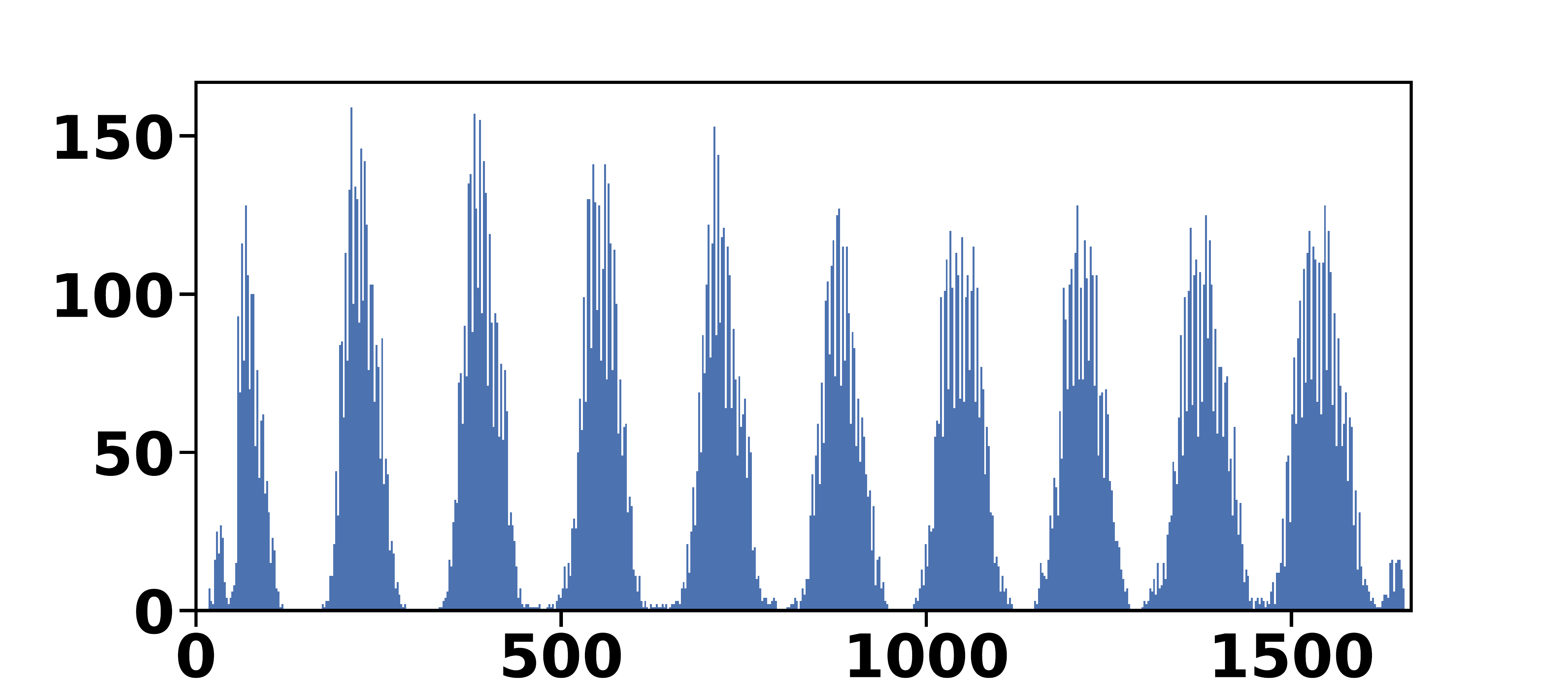}
		\vspace{-5.50mm}
		\caption{Pair 4}
	\end{subfigure}
	\begin{subfigure}[t]{0.3\columnwidth}
		\centering
		\includegraphics[width=1\columnwidth]{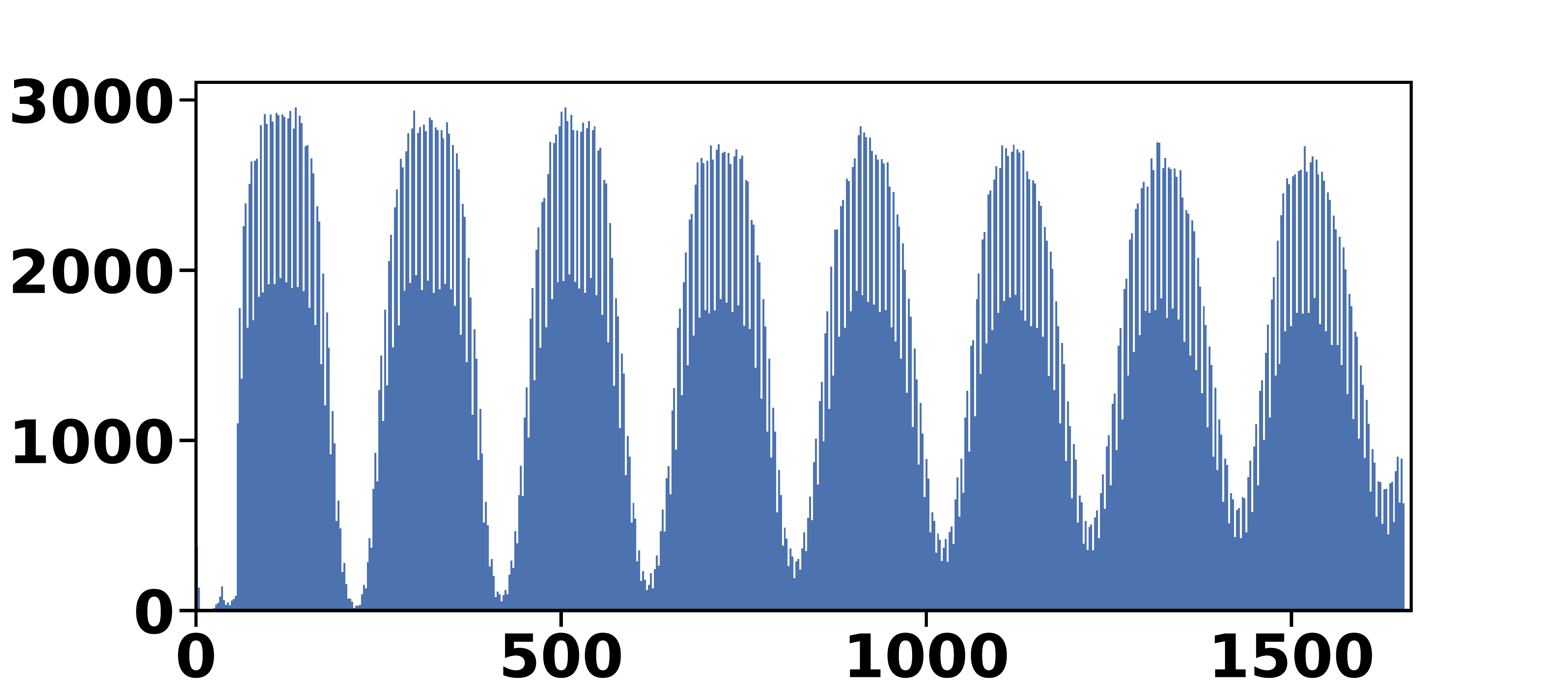}
		\vspace{-5.50mm}
		\caption{Pair 5}
		\label{fig:pair5}
	\end{subfigure}
	\begin{subfigure}[t]{0.3\columnwidth}
		\centering
		\includegraphics[width=1\columnwidth]{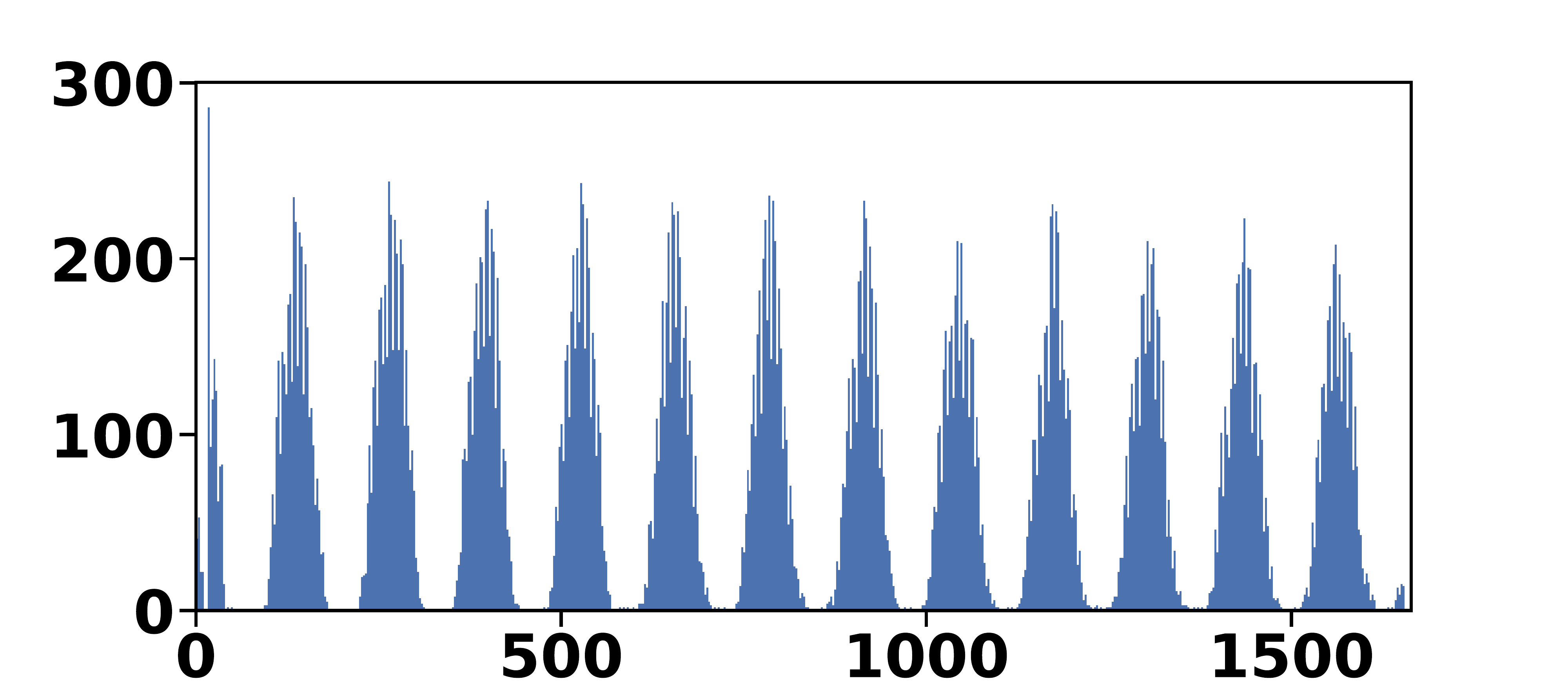}
		\vspace{-5.50mm}
		\caption{Pair 6}
	\end{subfigure}
	\begin{subfigure}[t]{0.3\columnwidth}
		\centering
		\includegraphics[width=1\columnwidth]{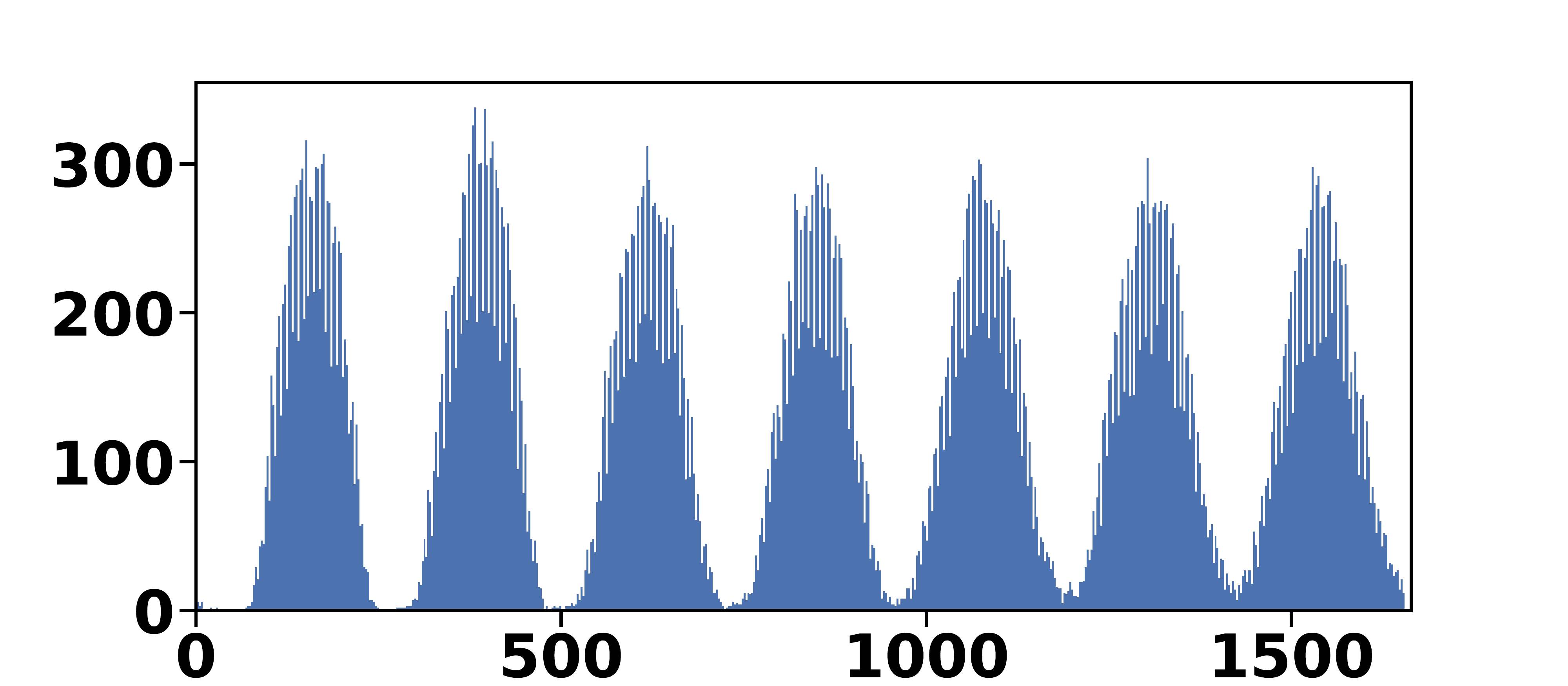}
		\vspace{-5.50mm}
		\caption{Pair 7}
	\end{subfigure}
	\begin{subfigure}[t]{0.3\columnwidth}
		\centering
		\includegraphics[width=1\columnwidth]{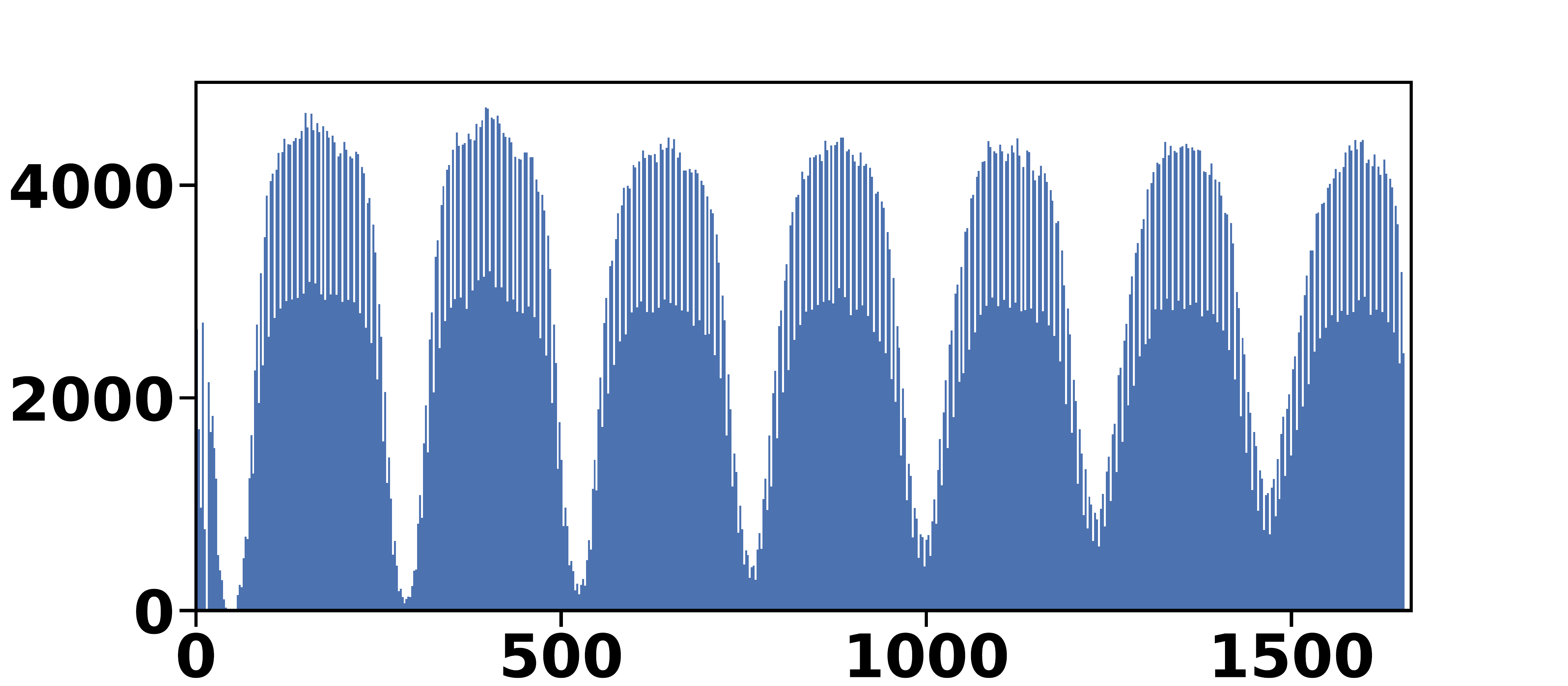}
		\vspace{-5.50mm}
		\caption{Pair 8}
		\label{fig:pair8}
	\end{subfigure}
	\begin{subfigure}[t]{0.3\columnwidth}
		\centering
		\includegraphics[width=1\columnwidth]{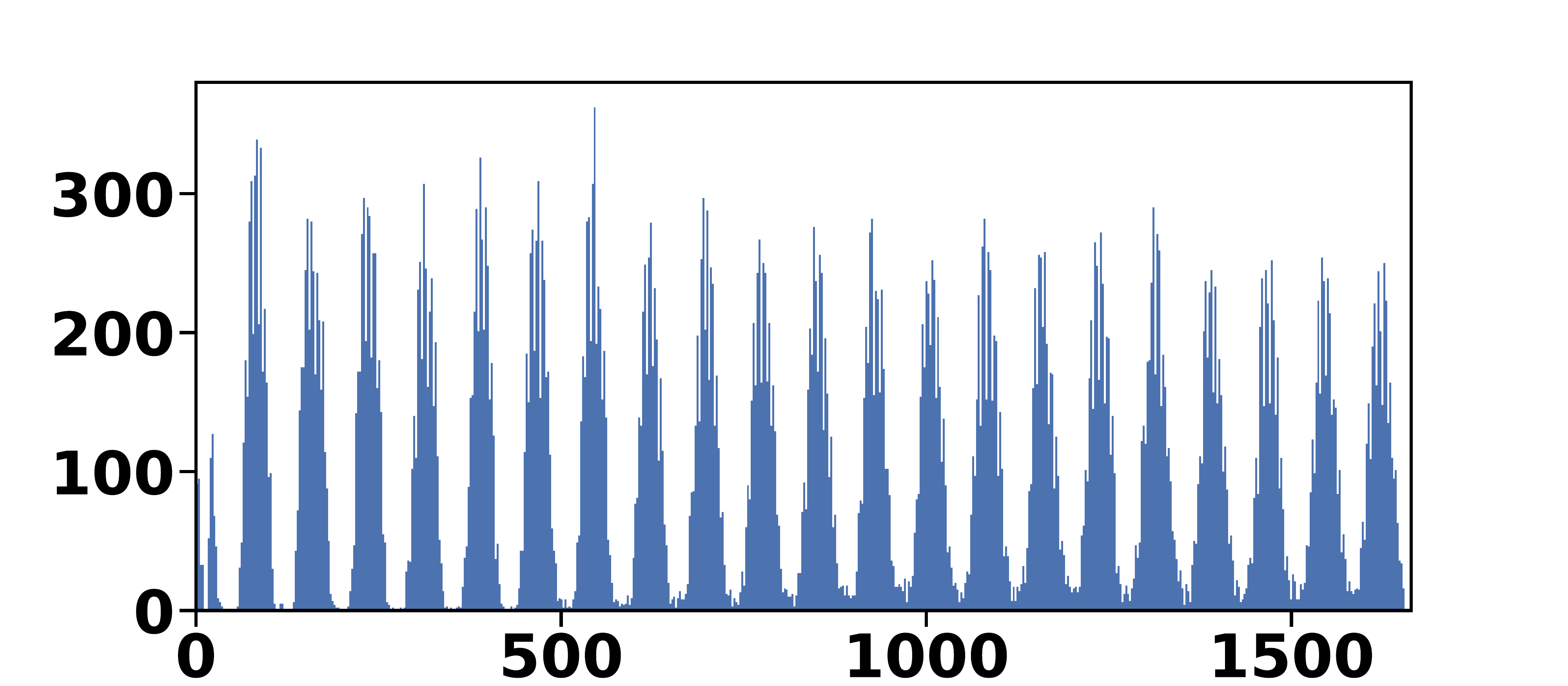}
		\vspace{-5.50mm}
		\caption{Pair 9}
	\end{subfigure}
	\begin{subfigure}[t]{0.5\columnwidth}
		\centering
		\vspace{+1.00mm}
		\includegraphics[width=1\columnwidth]{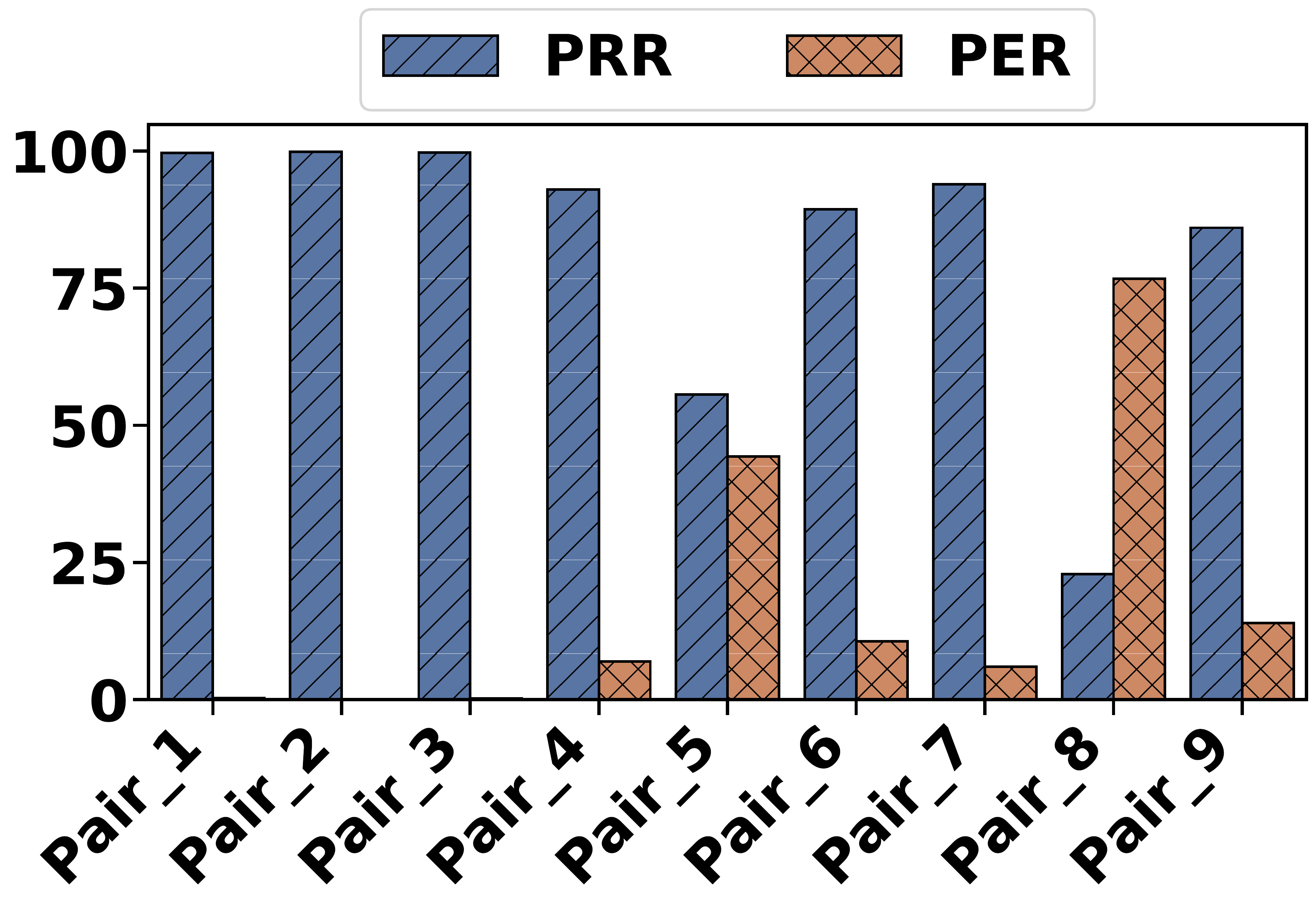}
		\vspace{-5.75mm}
		\caption{PRR and PER for all pairs}
		\label{fig:tosh_layoutcomp_prr}
	\end{subfigure}
	\vspace{-1.00mm}
	\caption{Error distribution across nine CT2 pairs when using the \blefive500K PHY (a--i) and overall CT performance (j).}
	\label{fig:tosh_layoutcomp}
	\vspace{-3.50mm}
\end{figure}

%---------------------------------------------------------%
\boldpar{Increasing CT density}
The effect of increasing CT density is explored in Fig.~\ref{fig:tosh_results_ct_comp}. Experiments were run across all PHYs for a single transmitter (\emph{no CT}), as well as increasing CT density from CT2 to CT12. Plots represent an average of multiple experiments run with randomly selected CT forwarders per experiment, while the same pseudo-random forwarding set remains consistent across each PHY. This averaging eliminates bias due to \emph{narrow and strong} beating experienced by CT2 pairs such as Figs.~\ref{fig:pair5}~and~\ref{fig:pair8}. Reliability drops significantly at CT3 due to the high data rate of the \blefive2M PHY, which requires a significant difference in received power between signals to experience the capture effect. This is consistent with recent literature~\cite{alnahas2020blueflood} and with our analysis in Sect.~\ref{sec:simulation}. While the \ieee PHY (on which most relevant CT literature is based) still performs well at CT3, its PRR also drops significantly at CT4. Interestingly, the \blefive1M PHY shows a gradual drop in performance at mid-level CT densities, while both \blefive uncoded PHYs (2M and 1M) experience a PRR `rally' at high CT density. We hypothesize that this is due to the increased diversity (i.e., additional paths and better chance of capturing dominant signals), or to the additional CT converging around an average RFO and spreading the effects of beating. %, and suggest this as an area for future research.
%\mb{we need to address why the PLR increases for coded PHYs. Is it due to the long preambles being lost in all the noise?}
%MB: I'll try and address this tomorrow (Friday)

%---------------------------------------------------------%
\boldpar{Beating effect across different CT pairs}
Fig.~\ref{fig:tosh_layoutcomp} examines the bit error distribution for nine different CT pairs on the \blefive500K PHY, chosen due to its beating sensitivity. 
%While the \blefive 500K PHY preamble is encoded at the same rate as 125K, the payload doesn't benefit from the same additional redundancy. 
The absence of the Manchester pattern mapper (as used in the 125K PHY) results in a high degree of corruption across the packet, meaning that beating errors are more prominent. Beating is therefore clearly seen across almost all pairs, with the exception of Figs.~\ref{fig:tosh_layoutcomp_1}~and~\ref{fig:tosh_layoutcomp_2}, where the RFO was not significant enough to result in observable beating (i.e., these pairs experience wide beating greater than the packet's transmission period). Fig.~\ref{fig:tosh_layoutcomp_prr} provides additional information about the PRR and PER for each pair, further showing how the RFO between pairs affects the beating width and, consequently, the performance of CT.

%---------------------------------------------------------%
%\subsubsection{Response to Different Data Patterns}
% TBD (extension?)

%---------------------------------------------------------%
%\subsubsection{Long-Term RFO Behavior (24H Experiments)}
% TBD (extension?)

%---------------------------------------------------------%
% Observations
%---------------------------------------------------------%
\subsection{Key Observations}
Recent literature has theorized that the beating effect should have a significant impact on CT performance. % have been inciteful. 
The experimental results presented in this section have shown that beating is present in both coded and uncoded \blefive PHYs, as well as in the DSSS-based \ieee PHY. We summarize these results by outlining a number of key observations.

\boldpar{Beating frequencies are device-specific} As shown by Fig.~\ref{fig:tosh_layoutcomp}, beating frequencies depend on the RFO between device pairs, and one cannot directly extrapolate results from a specific pair.

\boldpar{Preambles are sensitive to beating} While the start of a preamble can randomly coincide with a beating valley or peak, these results are relative to \emph{received} packets (i.e., those for which the preamble was successfully detected) and hence have bias towards a certain initial phase relationship.
This bias is further increased by calibration the receiver performs during the reception of the preamble~\cite{ble5specs}. As the packet is being received, the beating changes the signal properties and this calibration is no longer optimal; before periodically returning to the optimal operation point with a frequency equal to the beating frequency.  It is worth noting that this explains why beating is visible through an error distribution analysis, and that with no bias (i.e., no preamble) it would present as a flat error distribution.

\boldpar{High data rate PHYs benefit from packet repetition} As shown by Fig.~\ref{fig:tosh_phycomp}, packet transmissions in high data rate PHYs span fewer beating valleys (potentially zero if the packet period is shorter than the beating period). Since the position of peaks and valleys is random, after several repetitions, i.e., with a higher \texttt{TX\_N}, it is likely that a packet will not experience a valley during the preamble and will be correctly received. Note that \texttt{TX\_N} is a core component of many CT-based protocols.

\boldpar{Low data rate PHYs benefit from the coding gain} Fig.~\ref{fig:tosh_phycomp_ratio} shows a significant difference in reliability between the \blefive500K and 125K\,PHYs. Indeed, \blefive500K exhibits the worst performance of any of the PHYs compared in this section. It is likely that the convolutional coding employed by \blefive is sensitive to beating errors, while the gain seen using the 125K PHY stems from the additional pattern mapper redundancy over the payload (as mentioned in Sect.~\ref{sec:simulation}). Similarly, while not achieving the same gains of \blefive 125K PHY, the DSSS used in \ieee halves the PER in comparison to the \blefive500K PHY. It is worth noting that results in this section do not consider significant external noise or interference, which may particularly penalize the very long packets of the \blefive125K PHY, as seen in the following section, and the use of the 500K PHY may again constitute an effective trade-off in harsh environments, particularly to survive intermittent jamming.

\begin{figure}[!t]
	\centering
	\vspace{-1.75mm}
	\includegraphics[width=0.99\columnwidth]{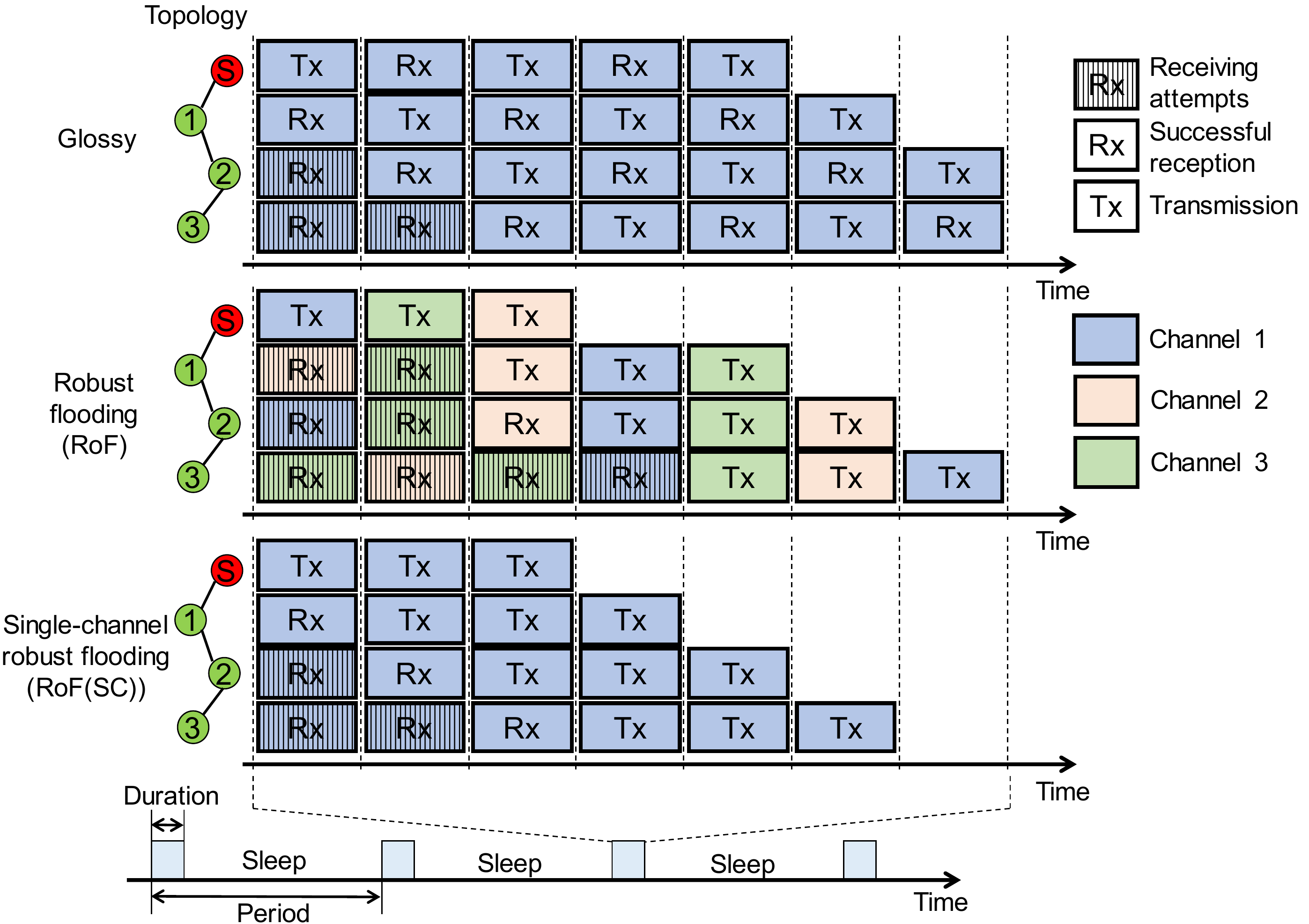}
	\vspace{-4.25mm}
	\caption{Operation of CT-based flooding primitives: Glossy, RoF, and RoF~(SC). In each case, the maximum number of transmissions~(\texttt{TX\_N}) is set to three.}
	\vspace{-0.25mm}
	\label{fig:protocols}
\end{figure}

\section{CT Performance over different PHYs: Experimental Evaluation with RF Interference}
\label{sec:interference}
%Unlike the previous section that evaluates the performance of single link between multiple transmitters and a single receiver, this section goes one step further from link layer to network-wide flooding protocols.
%We present an experimental study on the impact of different physical layers on CT-based protocols in the presence of RF interference.
This section presents an experimental study on the impact of different physical layers on CT-based protocols in the presence of RF interference.
Specifically, we evaluate  three CT-based flooding protocols on a large multi-hop network: Glossy~\cite{ferrari2011efficient}, Robust~Flooding~(RoF)~\cite{lim2017competition}, and Robust~Flooding~Single Channel~(RoF~(SC)), whose operations are depicted in Fig.~\ref{fig:protocols}.

Early CT literature adopted the flooding approach popularized in Glossy~\cite{ferrari2011efficient}, which triggers transmissions after successful receptions, thereby alternating \emph{Rx} and \emph{Tx} slots at each hop.
However, not only does the original Glossy approach operate on a single channel, meaning it is susceptible to RF interference at that frequency, but this reception-triggered \emph{Rx-Tx} technique means that \emph{Rx} failures will result in a missed transmission opportunity~\cite{lim2017competition, ma2018competition}. That is, using this technique, it is difficult to resume a CT flood if it is interrupted by interference.
An alternative approach was taken by the authors of~\cite{lim2017competition, raza2017competition}, introducing \emph{back-to-back} CT transmission slots (i.e.~\emph{Rx-Tx-Tx}) alongside robust frequency diversity through per-slot channel hopping. In this Robust Flooding (RoF) approach, the first transmission of a node is still triggered by correct reception, but further transmissions are time-triggered, with nodes synchronously hopping frequency at each slot. Specifically, at each slot, a node in RoF chooses a channel to from a successful reception channel list to receive.

We compare these two approaches (Glossy and RoF) as they are commonly used as primitives to construct more complex CT-based protocols, and are hence representative of wider CT literature. Furthermore, we introduce a variant of RoF -- the RoF Single Channel (RoF (SC)) -- to observe how this protocol performs w.r.t the single-channel environment used by Glossy.

%---------------------------------------------------------%
\subsection{Experimental Setup}

We evaluate each protocol (Glossy, RoF, and RoF (SC)) by computing three key dependability metrics: end-to-end \emph{reliability}, \emph{latency}, and \emph{energy consumption} -- all metrics measured in hardware by the \dcubee testbed.
We consider three scenarios characterized by the absence or presence of interference, denoted as \textbf{no}, \textbf{mild}, and \textbf{strong} interference.

\dcubee's controllable RF interference is generated by its observer nodes (Raspberry\,Pi\,3) using \jamlabng~\cite{schuss19jamlabng}.
\emph{Mild interference} (aka level 2 in \dcubee) uses a power of 30\,mW, generating interference for $\approx$\,5\,ms every 13\,ms period.
\emph{Strong interference} (aka level 3 in \dcubee) emulates the transmissions of multiple \wifi devices across all the 2.4\,GHz band. Each Raspberry Pi 3 node chooses a different channel, generating interference with a power of 200\,mW for $\approx$\,8\,ms every 13\,ms.

Each protocol (Glossy, RoF, and RoF (SC)) is run on \dcubee's data dissemination scenarios, i.e., those sending data from a single source to multiple destinations over a multi-hop network. To emulate event-based scenarios, we configure \dcubee to generate aperiodic messages with short (8B) payload for an alarm scenario and long (64B) payload for a condition monitoring scenario.
Other experimental parameters are set as follows.
We set the maximum number of transmission attempts per node during a flooding period (defined as \texttt{TX\_N} and set to three in the example shown in Fig.~\ref{fig:protocols}) to 6 for all protocols and fix the flooding periodicity to 200\,ms.
In Glossy and RoF (SC) the radio frequency is set to 2.480~GHz (i.e., channel 39 in BLE and channel 26 in \ieee), while RoF hops between 3 different channels (2.4025~GHz, 2.425~GHz, and 2.480~GHz). Finally, we set the transmission power to 0~dBm, which leads to a network diameter between 6 and 10 hops depending on the employed \dcubee layout. % the previous section used a transmission power of -8~dBm in order to more easily capture errors, these results use a more representative transmission power of 0~dBm.

All the results shown in this section utilize publicly available implementations of Glossy\footnote{https://github.com/ETHZ-TEC/Baloo/tree/2.0 - \emph{212dcde}} and RoF\footnote{https://github.com/ETHZ-TEC/robust-flooding - \emph{a03f38a}}, which we subsequently ported to the nRF52840-DK platform supported by \dcubee.
%Altogether, we perform \highlight{XX} experiments, for a total of over \highlight{XX} transmissions.

\begin{figure}[!t]
	\centering
	\begin{subfigure}[t]{1\columnwidth}
		\centering
		\includegraphics[width=1.00\columnwidth]{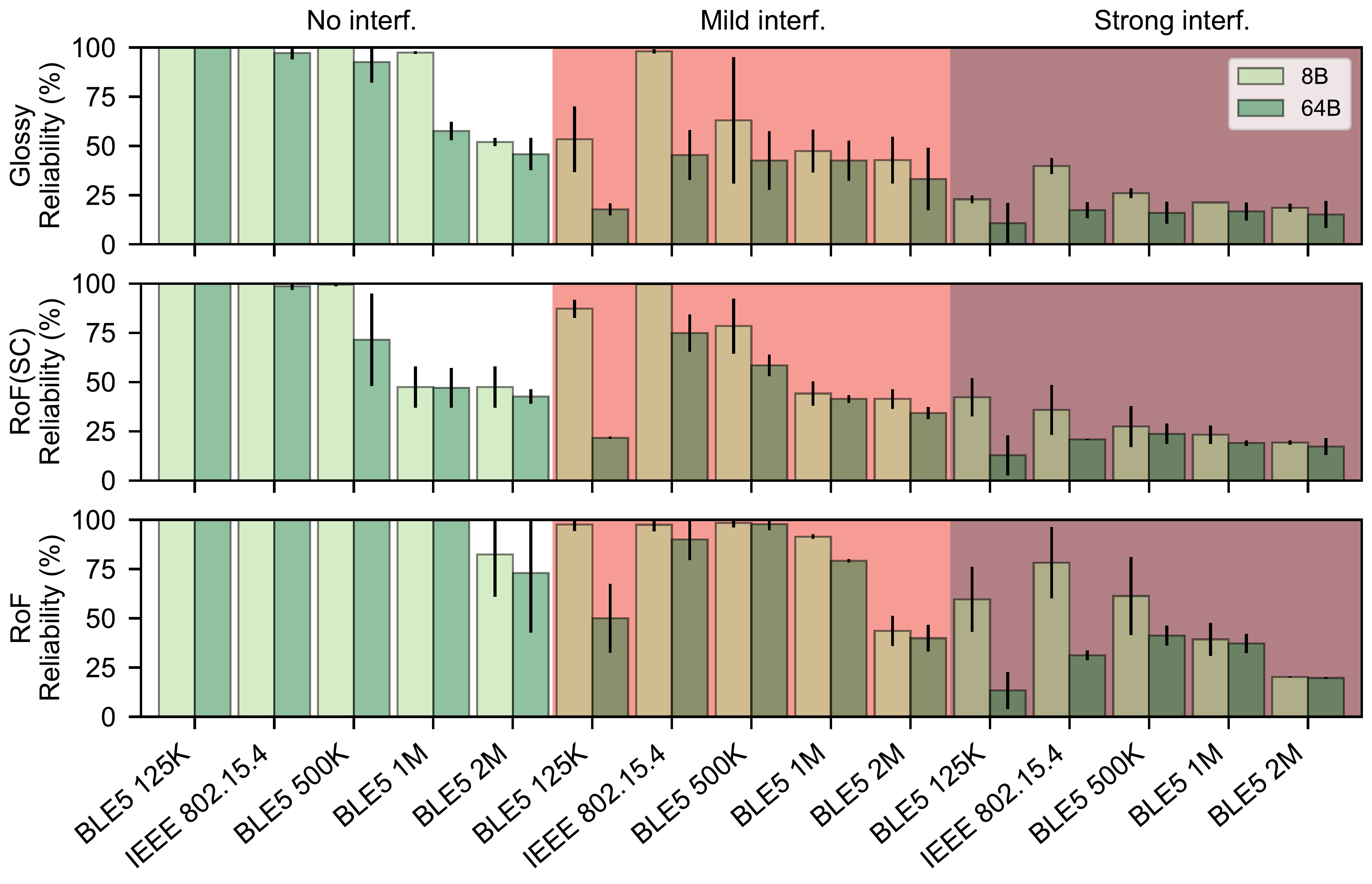}
		\vspace{-6.25mm}
		\caption{Average end-to-end reliability (\%).}
		\label{fig:reliability_0dbm}
	\end{subfigure}
	\begin{subfigure}[t]{1\columnwidth}
		\centering
		\vspace{0.4cm}
		\includegraphics[width=1.00\columnwidth]{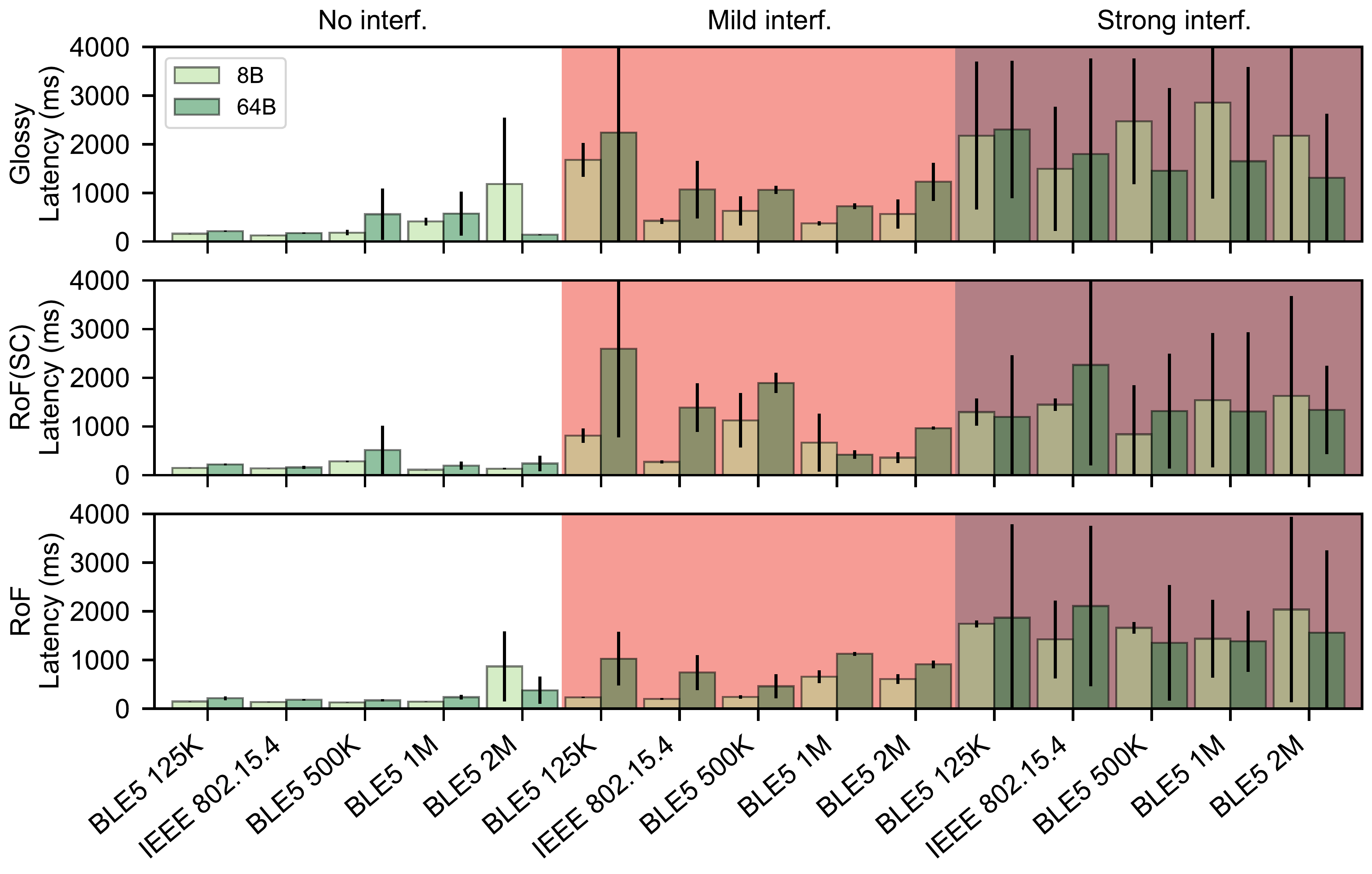}
		\vspace{-6.25mm}
		\caption{Average end-to-end latency (ms).}
		\label{fig:latency_0dbm}
		%		\vspace{0.1cm}
	\end{subfigure}
	\begin{subfigure}[t]{1\columnwidth}
		\centering
		\vspace{0.4cm}
		\includegraphics[width=1.00\columnwidth]{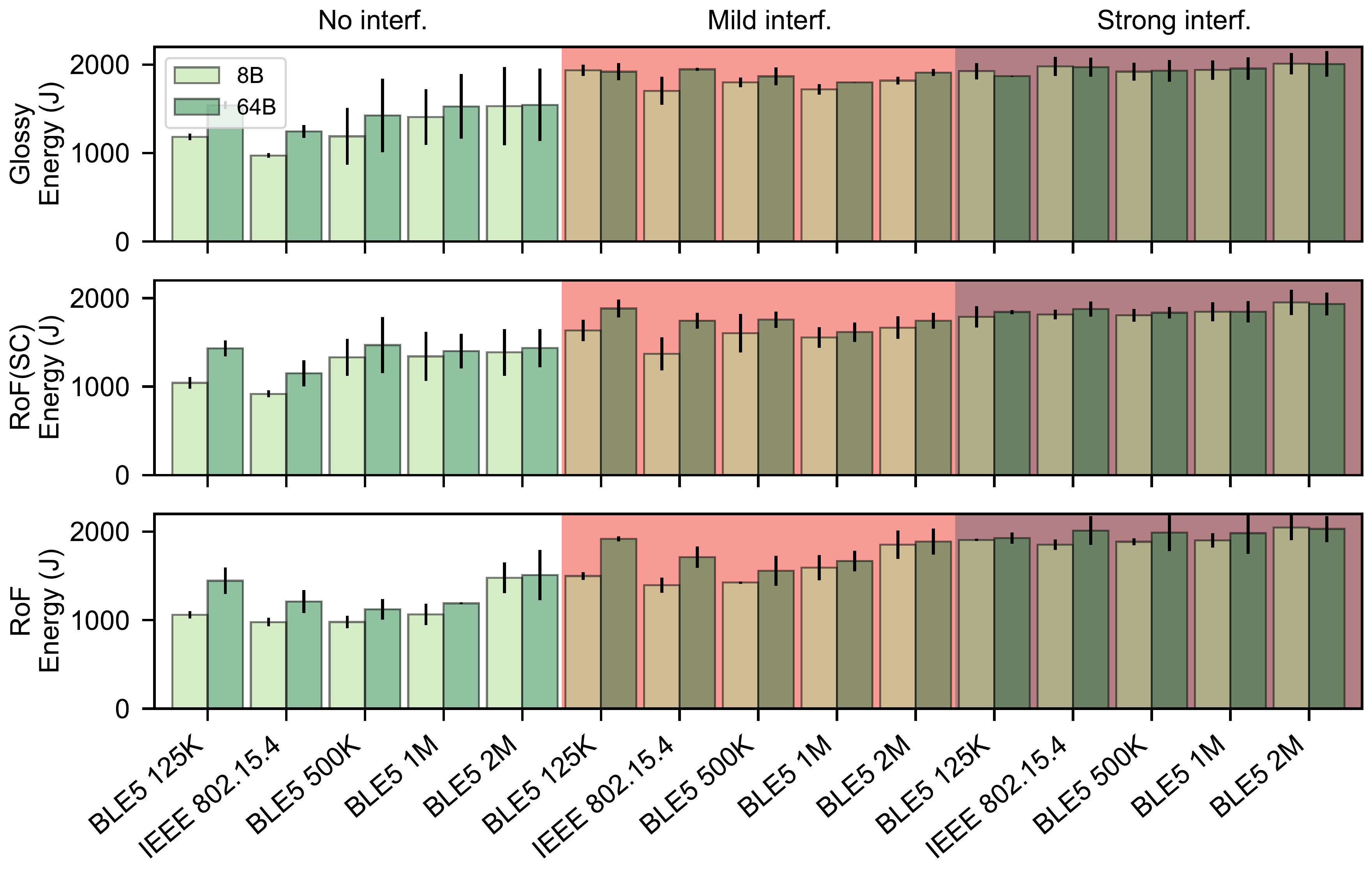}
		\vspace{-6.25mm}
		\caption{Total energy consumption (J).}
		\label{fig:energy_0dbm}
	\end{subfigure}
	\vspace{+1.00mm}
	\caption{Average reliability (a), latency (b), and total energy consumption (c) across all \dcubee data dissemination layouts.}
	\label{fig:protocol_results}
	\vspace{-3.00mm}
\end{figure}

\subsection{Results}
%---------------------------------------------------------%
\boldpar{Reliability}
% NO Inf. High data rate PHYs are bad due to lack of coding and beating
Prior to the introduction of external interference sources, Fig.~\ref{fig:reliability_0dbm} shows that, as data rate increases, the evaluated PHY layers struggle to maintain reliability.
%As previously mentioned in Sect.~\ref{sec:simulation} and Sect.~\ref{sec:beating}, not only do the \blefive high data rate PHYs lack redundancy to recover from errors, but they suffer from wider beating effects than their coded counterparts~\cite{liao2016revisiting}.
% NO Inf. RoF and RoF can be subject to desync due to drift. This will be worse at 2M (needs 0.25us sync rather than 0.5 at 1M)!
RoF exhibits the highest reliability out of the three evaluated protocols, while its single channel alternative, RoF (SC), decreases rapidly at higher data rates in comparison to Glossy.
Since the time-triggered \emph{back-to-back} transmission approach of RoF and RoF(SC) does not allow nodes to resynchronize at every \emph{Rx} slot (as in Glossy), nodes can be subject to synchronization errors due to drift. The high data rate PHYs are particularly sensitive to such errors: \blefive 2M is only able to tolerate CT synchronization errors of up to 0.25\,$\mu$s\cite{alnahas2019concurrentBLE5}.
% ALL. Short messages are better
In general, longer transmission times result in a greater chance of encountering interference and corrupting the packet. This is reflected in the reliability difference between \dcubee's long (64B) and short (8B) payloads under all three interference scenarios. This is additionally seen in the reliability of \blefive125K, where longer transmission times mean the PHY struggles to escape interference and results in surprisingly poor reliability across all three protocols.
% MILD and STRONG Inf. Time-triggered RoF is better
We also observe that the \emph{back-to-back} repetition of packets in RoF and RoF (SC) improves reliability over Glossy under mild and strong interference.
Furthermore, as expected, the frequency diversity introduced through RoF's channel hopping mechanism has a significant impact on the performance of all PHYs, except for the \blefive 2M PHY.  This is likely due to the higher data rate of this physical layer. As the interference generated by \jamlabng is periodic across multiple channels, if all RoF hopping channels are  occupied for the duration of that interference, then the flood will fail. However, the longer transmission time of the PHYs using a lower data rate practically increases the chance that one of the repeated transmissions is successful, hence allowing a node to escape interference.

% ALL. Sparse areas will cause beating errors - MB: Not sure we have anything to back this up!!!
%While we do not know the CT density at each node (due to the uncertain RF environment), \dcubee exhibits both sparse and dense areas.
%Sect.~\ref{sec:simulation} and Sect.~\ref{sec:beating} showed that in the sparser areas of the network significant beating effects could come into play if there is relatively little received power delta between CTs, and there will be a greater chance of hitting a beating valley over the longer on-air packet duration.

%---------------------------------------------------------%
\vspace{-1mm}
\boldpar{Latency}
% ALL. Interfernce icnreases latency. Not sure this is groundbreaking, but fine :P
The end-to-end latency of CT-based flooding protocols is inherently linked to their reliability. Indeed, the brute-force repetition inherent in CT-based flooding protocols means that packets \emph{may} successfully be received on poor channels, but much later in the flood.
Fig~\ref{fig:latency_0dbm} supports this, and we observe significant latency jumps as the amount of interference increases.
% ALL. High data rates phys with terrible reliability can have lowest latency
As \dcubee latency is only computed based on \emph{received} messages, it is conceivable that low latencies can be achieved even when reliability is poor. This is particularly apparent in \textbf{mild} interference, where we observe in Fig.~\ref{fig:latency_0dbm} that the \blefive uncoded PHYs exhibit low latency despite poor reliability.
In general, coded PHYs (i.e., \blefive125K, \blefive500K, and \ieee) have better performance with respect to latency.
% NO and MILD Inf. RoF (SC) 500K shows higher latencies
It is notable that under \textbf{no} and \textbf{mild} interference in RoF (SC), \blefive500K latency increases in comparison to other PHYs and the other two protocols (Glossy and RoF).
This is likely due to a combination of the lower data rate in comparison to the uncoded PHYs, alongside lower reliability due to its sensitivity to beating (as demonstrated in Sect.~\ref{sec:simulation}~and~\ref{sec:beating}).
% ALL. 125K shows low latency due to long range - so fewer hops
Finally, Fig.~\ref{fig:latency_0dbm} shows that, for shorter packets, the \blefive125K PHY enjoys low latency similar to the other PHYs. This is likely due to its improved receiver sensitivity, which provides longer transmission ranges (and, hence, the ability to span the network in fewer hops).

%---------------------------------------------------------%
\vspace{-1mm}
\boldpar{Energy}
% ALL. Poor reliability increases energy %
Similar to latency, the energy consumption of all nodes in the network is intrinsically linked to the overall reliability. Unsuccessful reception means the radio needs to remain on for a longer time. As shown in Fig.~\ref{fig:energy_0dbm}, although in principle higher data rate PHYs should have a lower energy consumption, this relationship with reliability means that for \emph{short} payloads \blefive1M and 2M are less energy efficient than the coded PHYs.
% ALL. PHY rates still matter. Low data rates with long packets will still incur large energy consumption.
However, the underlying PHY rate is still a fundamental factor in the node's energy consumption. For \emph{long} payloads, indeed, the lower data rates supported by the coded PHYs mean the radio can take a considerable time to transmit a packet, and incur greater energy consumption over the uncoded PHYs.

\vspace{-1mm}
\subsection{Key Observations}
%\cb{I think a "Key Observations" subsection summarizing the takeaway message is needed here. Observations should reflect the role of the beating effects, interesting performance of PHYs under interference, how the PHYs could be used to improve performance of CT-based protocols (e.g., what you used as a summary during the EWSN talk).}
With respect to these results, we make a number of key observations on the network-wide performance of CT-based flooding protocols under interference as a function of the employed PHY.

%MB: As stated above, I don't think we have the figures for sparese/dense networks now.
%\boldpar{Sparse networks are more susceptible to the beating effect}
%The amplitudes of valleys in the beating effect are small in sparse networks, which means it is more difficult for a receiver to decode these valleys.
\vspace{-1mm}
\boldpar{At a network level, high data rate PHYs struggle even under no interference}
Without the redundancy gains of coded PHYs, high data rate PHYs are sensitive to both de-synchronization and beating, particularly at greater CT density. Even with the added benefit of frequency diversity, RoF still cannot achieve high reliability when using the \blefive2M PHY.

%MB: As stated above, I don't think we have the figures for sparese/dense networks now.
%\boldpar{BLE 1 Mbps is the best in dense networks when interference is not strong}
%In dense networks,\blefive1M could achieve high reliability, low latency, and high energy efficiency if interference is just mild.\mb{Can we still say something about BLE1Mbps in general?}
\vspace{-1mm}
\boldpar{The \blefive125K PHY is not necessarily the answer}
Although performing well when there is no interference, \blefive125K suffers from poor performance as soon as interference kicks in and packet size increases, taking a relatively large hit with respect to latency and reliability in comparison to the other PHYs.

% MB: Probably should combine these two? Essentially they say we should use \ieee and \blefive 500K when there is ANY interference
%\boldpar{The PHYs of \ieee and \blefive 500K should be considered in sparse networks when interference is not strong} In sparse networks, PHYs of \ieee and \blefive 500K can achieve the highest reliability whilst have low latency and low energy consumption.
\vspace{-1mm}
\boldpar{\blefive500K and \ieee perform well under interference}
Under interference \blefive500K achieves higher reliability and similar or lower latency compared to other \blefive PHYs, while performing worse than the other coded PHYs when there is no interference.
This is consistent with the findings in Sect.~\ref{sec:simulation}, which showed that 500K will outperform other PHYs when there is a significant received power delta, which is likely to occur in high noise conditions.
Furthermore, \ieee, on which much of the CT literature is based, demonstrates similar high reliability under interference while benefiting from the increased data rate.
If the level of network interference is unknown, CT protocols benefit from transmission on either the \blefive 500K or \ieee PHYs.

\boldpar{RoF's time-triggered transmission and channel hopping produce significant gains}
It is clear from Fig.~\ref{fig:protocol_results} that the combination of \emph{back-to-back} time triggered transmissions and channel hopping employed in RoF provides significant gains under all interference scenarios.
However, results from RoF (SC) show that, without frequency diversity, there is a chance that at higher data rates, the interference duration may be longer than the flooding period. As a blunt instrument, \texttt{TX\_N} could therefore be increased to take advantage of greater temporal redundancy and improve protocol reliability under interference.

% Related Work
\vspace{-1.0mm}
\section{Related Work} \label{sec:related_work}
We discuss next related work and highlight how the contributions presented in Sect.~\ref{sec:simulation}~--~\ref{sec:interference} advance the state-of-the-art.

\vspace{-1mm}
\boldpar{CT on different PHYs}
After the influential work by Ferrari et al.~\cite{ferrari2011efficient} was published in 2011, a large number of researchers has started to study CT and develop CT-based protocols~\cite{zimmerling20synchronous, ferrari2012low, suzuki13choco, istomin2016data, doddavenkatappa2013splash, du17pando, landsiedel13chaos, chang18constructive}.
While most of the early works targeted exclusively \ieee devices using the 2.4\,GHz band, in the last years, a few studies have shown the feasibility of CT on other physical layers supported by \ieee, such as the UWB PHY~\cite{kempke16surepoint, corbalan19secon, lobba20concurrent}, as well as sub-GHz short-range~\cite{liao16toward, beutel19ipsn} and long-range technologies~\cite{liao17lora, ma20chirpbox}.

A number of works have recently focused on studying the feasibility of CT on BLE~\cite{alnahas2019concurrentBLE5, schaper2019truth, alnahas2020blueflood, roest15ble}.
Specifically, Al Nahas et al.~\cite{alnahas2019concurrentBLE5} verified the feasibility of a CT-based flooding protocol, named BlueFlood, on the different \blefive PHYs experimentally, and also reported its performance on \ieee~\cite{alnahas2020blueflood}.
Schaper~\cite{schaper2019truth} studied the conditions to make CT successful in these PHYs in an anechoic chamber.

Different from these studies, our current work does not aim to prove the \textit{feasibility} of CT on different radio technologies or PHYs, but instead to provide an in-depth \textit{characterization of the role of the physical layer} on the reliability and efficiency of CT-based solutions employing \ieee and \blefive.
To the best of our knowledge, we are the first do this in a systematic manner by demonstrating experimentally the impact of errors induced by de-synchronization and beating distortion in CT-based protocols as a function of the employed PHY.

\vspace{-1mm}
\boldpar{CT performance under interference}
Several works have shown that CT-based data collection and dissemination protocols can outperform conventional routing-based solutions in terms of reliability, end-to-end, and energy consumption even in the presence of harsh radio interference~\cite{lim2017competition, istomin18crystal, ma20harmony, escobar2019competition}, as also highlighted in the context of the EWSN dependability competition series~\cite{boano17competition, schuss17competition, schuss18benchmark}.
To sustain a dependable performance under interference, CT-based solutions have been enriched, among others, with mechanisms such as local opportunistic retransmissions~\cite{ferrari2011efficient, sutton_zippy_2015, suzuki15ewsn}, channel-hopping~\cite{lim2017competition, istomin18crystal, sommer16channelhopping, escobar2016redfixhop, doddavenkatappa14p3}, network coding~\cite{doddavenkatappa2013splash, du17pando, yuan15ripple, mohammad18codecast, herrmann18mixer}, noise detection~\cite{istomin18crystal, ma20harmony}, stretched preambles~\cite{escobar2019competition}, data freezing~\cite{ma20harmony}, as well as an improved understanding of the network state~\cite{carlson13forwarder, brachmann16laneflood, sarkar16sleepingbeauty}.

However, most of these protocols have been implemented for and evaluated with \ieee technology only.
In this paper, we are the first to study the performance of CT-based data collection protocols on a large scale under interference \textit{as a function of the employed PHY}.
We did this by evaluating the dependability of CT-based protocols on a large scale using modern multi-radio platforms supporting several PHYs, and by analyzing the impact of other factors such as the length of the transmitted messages and the harshness of the interference.

\vspace{-1mm}
\boldpar{Impact of beating effect on CT}
A few studies have tried to underpin the foundations of concurrent transmissions on a signal level.
Ferrari et al.~\cite{ferrari2011efficient} have simulated CT signals with Matlab and explained how accurately packets should be aligned in order to design reliable protocols.
Other works~\cite{rao_murphy_2016, wang_triggercast_2013, wang_disco:_2015} have analyzed CT signals theoretically and argued that it is difficult to generate ideal constructive interference, due to the timing errors caused by radio propagation and clock drift.

Liao et al. have been the first to argue that there exists a beating effect caused by innate CFO between device oscillators~\cite{liao2016revisiting}.
Specifically, in~\cite{liao2016revisiting}, the resultant signals are generated by Matlab and a TelosB node was used to observe how the DSSS modulation in \ieee saves CT signals from the beating effect.
More recent studies have demonstrated these beating effects generate periods of both constructive and destructive interference by observing the raw IQ samples of devices connected to an SDR using coaxial cables~\cite{alnahas2020blueflood}.

In this paper, to the best of our knowledge, we are the first to demonstrate how physical layer effects such as CFO-induced beating, and de-synchronization from hardware clock drift, directly affect the signal observed by a receiving node \textit{using over-the-air testbed experiments}.

\section{Conclusions and Future Work} \label{sec:conclusions}
To date, a significant volume of work has shown that CT-based protocols have an important role to play in providing robust and low-latency communication in
%both single and multi-hop 
mesh networks. Particularly in high interference environments, CT is a valuable tool allowing designers of low-power wireless protocols to mitigate the impact of harsh RF conditions. This paper provides the first systematic experimental evaluation into the role of \ieee and \blefive PHY layers on CT performance, with important insights into how the choice of the physical layer can exacerbate or reduce errors due to beating, de-synchronization, and external radio interference.

Specifically, we find that the coding used by the \blefive500K PHY is effective against interference in sparse networks, but ineffective against beating, whereas the \blefive125K PHY is effective against beating, but long transmission times may mean that packets fail under intermittent interference conditions as modeled in \dcubee. Furthermore, we show that the \blefive1M~and~2M high data-rate PHYs are not particularly robust against interference, but that repetition and short packet on-air times can improve the overall performance against beating. We conclude that the \blefive500K or \ieee PHYs should be used for long packets when operating in harsh wireless conditions, whereas the \blefive125K PHY should be used for short packets and when the interference is not high or the number of CT is high enough to increase the SNR (dense networks). On the other hand, in low-noise environments with few CT, the use of the \blefive2M PHY reduces errors due to beating, while supporting far higher data-rates. As the number of CT increases, however, \blefive125K provides significant gains over other PHYs.

While these findings are important to the design of CT protocols, there are a number of key areas that require additional research and further clarification. Crucially, greater understanding is needed around how beating errors affect a protocol's scalability on a network level.
%one cannot draw direct parallels between the single-hop results presented in Sect.~\ref{sec:simulation}~and~\ref{sec:beating}, and the multi-hop scenarios presented in Sect.~\ref{sec:interference}. 
While Sect.~\ref{sec:beating} presents results on the impact of CT density, real-world RF conditions and deployments make it difficult to know the amount of concurrent transmissions received at each node. Furthermore, CFO is particularly sensitive to temperature. The relative frequency offset may therefore change over time, resulting in changes to the beating frequency. 

Finally, this paper has highlighted the effectiveness of coding in improving CT reliability. Techniques such as \emph{interleaving}, i.e., bit shuffling, (which improves the robustness of forward error correction with respect to burst errors) and \emph{bit voting} are missing in the analyzed physical layers, and would be a very effective addition to increase the reliability of CT.

% Depending on space and how much we want to give away, we may want to also talk about the following.
%TODO: simulations on intermittent interfernce (to more closely align with dcube as simulations are QWGN)
%TODO: Coding for 1M/2M, tailored to CT.
%TODO: Bit voting over repetitions.
%TODO: Single-hop and constant noise vs. multi-hop and intermitent jamming.
%TODO: Use the GPIO to show HF clock frequency.
%TODO: Does the choice of baseband freq/channel have any impact.
%TODO: Show 125K linear desync errors.
%TODO: Show change in beating over a 24H period.
%TODO: Add \ieee to simulations?

% Bibliography
%\Urlmuskip=0mu plus 1mu\relax
\balance
\bibliographystyle{unsrt}
\bibliography{arxiv}  % sigproc.bib is the name of the Bibliography in this case

\begin{thebibliography}{10}

\bibitem{leentvaar76capture}
Krijn Leentvaar and Jan~H. Flint.
\newblock {The Capture Effect in FM Receivers}.
\newblock {\em IEEE Transactions on Communications}, 24(5):531--539, May 1976.

\bibitem{liao2016revisiting}
Chun-Hao Liao, Yuki Katsumata, Makoto Suzuki, and Hiroyuki Morikawa.
\newblock {Revisiting the So-Called Constructive Interference in Concurrent
  Transmission}.
\newblock In {\em Proceedings of the $41^{st}$ International Conference on
  Local Computer Networks ({LCN})}, pages 280--288. IEEE, November 2016.

\bibitem{zimmerling20synchronous}
Marco Zimmerling, Luca Mottola, and Silvia Santini.
\newblock {Synchronous Transmissions in Low-Power Wireless: A Survey of
  Communication Protocols and Network Services}.
\newblock {\em CORR -- arXiv preprint 2001.08557}, January 2020.

\bibitem{ferrari2011efficient}
Federico Ferrari, Marco Zimmerling, Lothar Thiele, and Olga Saukh.
\newblock {Efficient Network Flooding and Time Synchronization with Glossy}.
\newblock In {\em Proceedings of the $10^{th}$ International Conference on
  Information Processing in Sensor Networks ({IPSN})}, pages 73--84. IEEE,
  April 2011.

\bibitem{ferrari2012low}
Federico Ferrari, Marco Zimmerling, Luca Mottola, and Lothar Thiele.
\newblock {Low-Power Wireless Bus}.
\newblock In {\em Proceedings of the $10^{th}$ International Conference on
  Embedded Network Sensor Systems ({SenSys})}, pages 1--14. ACM, November 2012.

\bibitem{suzuki13choco}
Makoto Suzuki, Yasuta Yamashita, and Hiroyuki Morikawa.
\newblock {Low-Power, End-to-End Reliable Collection Using Glossy for Wireless
  Sensor Networks}.
\newblock In {\em Proceedings of the $77^{th}$ International Vehicular
  Technology Conference ({VTC})}, pages 1--5. IEEE, June 2013.

\bibitem{istomin2016data}
Timofei Istomin, Amy~Lynn Murphy, Gian~Pietro Picco, and Usman Raza.
\newblock {Data Prediction + Synchronous Transmissions = Ultra-Low Power
  Wireless Sensor Networks}.
\newblock In {\em Proceedings of the $14^{th}$ International Conference on
  Embedded Network Sensor Systems ({SenSys})}, pages 83--95. ACM, November
  2016.

\bibitem{doddavenkatappa2013splash}
Manjunath Doddavenkatappa, Mun~Choon Chan, and Ben Leong.
\newblock {Splash: Fast Data Dissemination with Constructive Interference in
  Wireless Sensor Networks}.
\newblock In {\em Proceedings of the $10^{th}$ International Symposium on
  Networked Systems Design and Implementation ({NSDI})}, pages 269--282.
  USENIX, April 2013.

\bibitem{du17pando}
Wan Du, Jansen~Christian Liando, Huanle Zhang, and Mo~Li.
\newblock {Pando: Fountain-Enabled Fast Data Dissemination With Constructive
  Interference}.
\newblock {\em IEEE/ACM Transactions on Networking}, 25(2):820--833, April
  2017.

\bibitem{lim2017competition}
Roman Lim, Reto Da~Forno, Felix Sutton, and Lothar Thiele.
\newblock {Competition: Robust Flooding using Back-to-Back Synchronous
  Transmissions with Channel-Hopping}.
\newblock In {\em Proceedings of the $14^{th}$ International Conference on
  Embedded Wireless Systems and Networks ({EWSN}), competition session}, pages
  270--271. Junction Publishing, February 2017.

\bibitem{istomin18crystal}
Timofei Istomin, Matteo Trobinger, Amy~Lynn Murphy, and Gian~Pietro Picco.
\newblock {Interference-Resilient Ultra-low Power Aperiodic Data Collection}.
\newblock In {\em Proceedings of the $17^{th}$ International Conference on
  Information Processing in Sensor Networks ({IPSN})}, pages 84--95. ACM, April
  2018.

\bibitem{ma20harmony}
Xiaoyuan Ma, Peilin Zhang, Ye~Liu, Carlo~Alberto Boano, Hyung-Sin Kim, Jianming
  Wei, and Jun Huang.
\newblock {Harmony: Saving Concurrent Transmissions from Harsh RF
  Interference}.
\newblock In {\em Proceedings of the $39^{th}$ International Conference on
  Computer Communication ({INFOCOM})}. IEEE, July 2020.

\bibitem{escobar2019competition}
Antonio Escobar-Molero, Javier Garcia-Jimenez, Jirka Klaue, Fernando
  Moreno-Cruz, Borja Saez, Francisco~J Cruz, Unai Ruiz, and Angel Corona.
\newblock {Competition: RedNodeBus, Stretching out the Preamble}.
\newblock In {\em Proceedings of the $16^{th}$ International Conference on
  Embedded Wireless Systems and Networks ({EWSN}), competition session}, pages
  304--305. Junction Publishing, February 2019.

\bibitem{boano17competition}
Carlo~Alberto Boano, Markus Schu{\ss}, and Kay R\"{o}mer.
\newblock {EWSN Dependability Competition: Experiences and Lessons Learned}.
\newblock {\em IEEE Internet of Things Newsletter}, March 2017.

\bibitem{polastre05telos}
Joseph Polastre, Robert Szewczyk, and David~E. Culler.
\newblock {Telos: Enabling Ultra-Low Power Wireless Research}.
\newblock In {\em Proceedings of the $4^{th}$ International Symposium on
  Information Processing in Sensor Networks ({IPSN})}, pages 364--369. IEEE,
  April 2005.

\bibitem{802154_ieee}
{IEEE 802.15.4 Working Group}.
\newblock {\em {IEEE Standard for Low-Rate Wireless Networks}}, {IEEE Std
  802.15.4-2015 (Revision of IEEE Std 802.15.4-2011, IEEE Std 802.15.4-2006,
  and IEEE Std 802.15.4-2003)} edition, April 2016.

\bibitem{wilhelm14concurrent}
Matthias Wilhelm, Vincent Lenders, and Jens~B. Schmitt.
\newblock {On the Reception of Concurrent Transmissions in Wireless Sensor
  Networks}.
\newblock {\em IEEE Transactions on Wireless Communications},
  13(12):6756--6767, August 2014.

\bibitem{alnahas2019concurrentBLE5}
Beshr~Al Nahas, Simon Duquennoy, and Olaf Landsiedel.
\newblock {Concurrent Transmissions for Multi-Hop Bluetooth 5}.
\newblock In {\em Proceedings of the $16^{th}$ International Conference on
  Embedded Wireless Systems and Networks ({EWSN})}, pages 130--141. Junction
  Publishing, February 2019.

\bibitem{schaper2019truth}
Anna-Brit Schaper.
\newblock {Truth be Told: Benchmarking BLE and IEEE 802.15.4}.
\newblock Master's thesis, ETH Zurich, Zurich, Switzerland, November 2019.

\bibitem{spoerk19ble5phy}
Michael Sp\"{o}rk, Carlo~Alberto Boano, and Kay R\"{o}mer.
\newblock {Performance and Trade-offs of the new PHY Modes of BLE 5}.
\newblock In {\em Proceedings of the International Workshop on Pervasive
  Systems in the IoT Era ({PERSIST-IoT})}, pages 7--12. ACM, July 2019.

\bibitem{cc2652_productsheet}
{Texas Instruments}.
\newblock {CC2652R SimpleLink Multiprotocol 2.4 GHz Wireless MCU datasheet,
  Rev. G}.
\newblock [Online] \url{https://www.ti.com/product/CC2652R} -- Last accessed:
  2020-05-26.

\bibitem{nrf52840_productsheet}
{Nordic Semiconductors}.
\newblock {nRF52840 Product Specification, v1.1}.
\newblock [Online]
  \url{https://infocenter.nordicsemi.com/pdf/nRF52840_PS_v1.1.pdf} -- Last
  accessed: 2020-05-26.

\bibitem{escobar19imprel}
Antonio Escobar-Molero.
\newblock {Improving Reliability and Latency of Wireless Sensor Networks Using
  Concurrent Transmissions}.
\newblock {\em at-Automatisierungstechnik}, 67(1):42--50, 2019.

\bibitem{schuss17competition}
Markus Schu{\ss}, Carlo~Alberto Boano, Manuel Weber, and Kay R\"{o}mer.
\newblock {A Competition to Push the Dependability of Low-Power Wireless
  Protocols to the Edge}.
\newblock In {\em Proceedings of the $14^{th}$ International Conference on
  Embedded Wireless Systems and Networks ({EWSN})}, pages 54--65. Junction
  Publishing, February 2017.

\bibitem{schuss18benchmark}
Markus Schu{\ss}, Carlo~Alberto Boano, and Kay R\"{o}mer.
\newblock {Moving Beyond Competitions: Extending D-Cube to Seamlessly Benchmark
  Low-Power Wireless Systems}.
\newblock In {\em Proceedings of the $1^{st}$ International Workshop on
  Benchmarking Cyber-Physical Networks and Systems ({CPSBench})}, pages 30--35.
  IEEE, April 2018.

\bibitem{schuss19jamlabng}
Markus Schu{\ss}, Carlo~Alberto Boano, Manuel Weber, Matthias Schulz, Matthias
  Hollick, and Kay R\"{o}mer.
\newblock {JamLab-NG: Benchmarking Low-Power Wireless Protocols under
  Controllable and Repeatable Wi-Fi Interference}.
\newblock In {\em Proceedings of the $16^{th}$ International Conference on
  Embedded Wireless Systems and Networks ({EWSN})}, pages 83--94. Junction
  Publishing, February 2019.

\bibitem{pasupathy1979minimum}
Subbarayan Pasupathy.
\newblock {Minimum Shift Keying: A Spectrally Efficient Modulation}.
\newblock {\em IEEE Communications Magazine}, 17(4):14--22, July 1979.

\bibitem{alnahas2020blueflood}
Beshr~Al Nahas, Antonio Escobar-Molero, Jirka Klaue, Simon Duquennoy, and Olaf
  Landsiedel.
\newblock {BlueFlood: Concurrent Transmissions for Multi-Hop Bluetooth
  5--Modeling and Evaluation}.
\newblock {\em CORR -- arXiv preprint 2002.12906}, February 2020.

\bibitem{escobar20phd}
Antonio Escobar-Molero.
\newblock {\em Using Concurrent Transmissions to Improve the Reliability and
  Latency of Low-Power Wireless Mesh Networks}.
\newblock Phd thesis, RWTH Aachen University, Aachen, Germany, June 2020.

\bibitem{viterbi67code}
Andrew~J. Viterbi.
\newblock {Error Bounds for Convolutional Codes and an Asymptotically Optimum
  Decoding Algorithm}.
\newblock {\em {IEEE Transactions on Information Theory}}, 13(2):260--269,
  April 1967.

\bibitem{ble5specs}
{Bluetooth Working Group}.
\newblock {Bluetooth Core Specification, Revision 5.2}, December 2019.

\bibitem{ble5distance}
{Bluetooth Blog}.
\newblock {Exploring Bluetooth 5 -- Going the Distance}.
\newblock [Online]
  \url{https://www.bluetooth.com/blog/exploring-bluetooth-5-going-the-distance/}
  -- Last accessed: 2020-05-26.

\bibitem{baddeley2019atomic}
Michael Baddeley, Usman Raza, Mahesh Sooriyabandara, George Oikonomou, Reza
  Nejabati, and Dimitra Simeonidou.
\newblock {Atomic-SDN: Is Synchronous Flooding the Solution to Software-Defined
  Networking in IoT?}
\newblock {\em {IEEE Access}}, 7:96019--96034, May 2019.

\bibitem{baddeley2020thesis}
Michael Baddeley.
\newblock {\em {Software Defined Networking for the Industrial Internet of
  Things}}.
\newblock PhD thesis, Dept. of Electrical and Electronic Engineering, Univ.
  Bristol, UK, 2020.

\bibitem{baddeley2019competition}
Michael Baddeley, Aleksandar Stanoev, Usman Raza, Yichao Jin, and Mahesh
  Sooriyabandara.
\newblock {Competition: Adaptive Software Defined Scheduling of Low Power
  Wireless Networks}.
\newblock In {\em Proceedings of the $16^{th}$ International Conference on
  Embedded Wireless Systems and Networks ({EWSN}), competition session}, pages
  298--299. Junction Publishing, February 2019.

\bibitem{ma2018competition}
Xiaoyuan Ma, Peilin Zhang, Weisheng Tang, Xin Li, Wangji He, Fuping Zhang,
  Jianming Wei, and Oliver Theel.
\newblock {Using Enhanced OFPCOIN to Monitor Multiple Concurrent Events under
  Adverse Conditions}.
\newblock In {\em Proceedings of the $15^{th}$ International Conference on
  Embedded Wireless Systems and Networks ({EWSN}), competition session}, pages
  211--212. Junction Publishing, February 2018.

\bibitem{raza2017competition}
Usman Raza, Yichao Jin, and Mahesh Sooriyabandara.
\newblock {Competition: Synchronous Transmissions based Flooding for Dependable
  Internet of Things.}
\newblock In {\em EWSN}, pages 278--279, 2017.

\bibitem{landsiedel13chaos}
Olaf Landsiedel, Federico Ferrari, and Marco Zimmerling.
\newblock Chaos: Versatile and efficient all-to-all data sharing and in-network
  processing at scale.
\newblock In {\em Proceedings of the $11^{th}$ International Conference on
  Embedded Networked Sensor Systems ({SenSys})}, pages 1--14. ACM, November
  2013.

\bibitem{chang18constructive}
Tengfei Chang, Thomas Watteyne, and Xavier Vilajosana Pedro~Henrique Gomes.
\newblock {Constructive Interference in 802.15.4: A Tutorial}.
\newblock {\em IEEE Communications Surveys and Tutorials}, 21(1):217--237,
  September 2018.

\bibitem{kempke16surepoint}
Benjamin Kempke, Pat Pannuto, Bradford Campbell, and Prabal Dutta.
\newblock {SurePoint: Exploiting Ultra Wideband Flooding and Diversity to
  Provide Robust, Scalable, High-Fidelity Indoor Localization}.
\newblock In {\em Proceedings of the $14^{th}$ {ACM} International Conference
  on Embedded Network Sensor Systems ({SenSys})}, pages 137--149. ACM, November
  2016.

\bibitem{corbalan19secon}
Davide Vecchia, Pablo Corbal\'{a}n, Timofei Istomin, and Gian~Pietro Picco.
\newblock {Playing with Fire: Exploring Concurrent Transmissions in
  Ultra-wideband Radios}.
\newblock In {\em Proceedings of the $18^{th}$ International Conference on
  Sensing, Communication and Networking ({SECON})}, pages 1--9. IEEE, June
  2019.

\bibitem{lobba20concurrent}
Diego Lobba, Matteo Trobinger, Davide Vecchia, Timofei Istomin, and Gian~Pietro
  Picco.
\newblock {Concurrent Transmissions for Multi-hop Communication on
  Ultra-wideband Radios}.
\newblock In {\em Proceedings of the $17^{th}$ International Conference on
  Embedded Wireless Systems and Networks ({EWSN})}, pages 132--143. Junction
  Publishing, February 2020.

\bibitem{liao16toward}
Chun-Hao Liao, Makoto Suzuki, and Hiroyuki Morikawa.
\newblock {Toward Robust Concurrent Transmission for Sub-GHz Non-DSSS
  Communication}.
\newblock In {\em Proceedings of the $14^{th}$ International Conference on
  Embedded Network Sensor Systems ({SenSys}), poster session}, pages 354--355.
  ACM, November 2016.

\bibitem{beutel19ipsn}
Jan Beutel, Roman Tr\"{u}b, Reto {Da Forno}, Markus Wegmann, Tonia Gsell,
  Romain Jacob, Michael Keller, Felix Sutton, and Lothar Thiele.
\newblock {The Dual Processor Platform Architecture}.
\newblock In {\em Proceedings of the $18^{th}$ International Conference on
  Information Processing in Sensor Networks ({IPSN}), demo session}, pages
  335--336. IEEE, April 2019.

\bibitem{liao17lora}
Chun-Hao Liao, Guibing Zhu, Daiki Kuwabara, Makoto Suzuki, and Hiroyuki
  Morikawa.
\newblock {Multi-Hop LoRa Networks Enabled by Concurrent Transmission}.
\newblock {\em {IEEE Access}}, 5:21430--21446, September 2017.

\bibitem{ma20chirpbox}
Xiaoyuan Ma, Dan Li, Fengxu Yang, Carlo~Alberto Boano, Pei Tian, and Jianming
  Wei.
\newblock {Chirpbox -- A Low-Cost LoRa Testbed Solution}.
\newblock In {\em Proceedings of the $17^{th}$ International Conference on
  Embedded Wireless Systems and Networks ({EWSN}), poster session}. Junction
  Publishing, February 2020.

\bibitem{roest15ble}
Coen Roest.
\newblock {Enabling the Chaos Networking Primitive on Bluetooth LE}.
\newblock Master's thesis, TU Delft, Delft, The Netherlands, October 2015.

\bibitem{sutton_zippy_2015}
Felix Sutton, Bernhard Buchli, Jan Beutel, and Lothar Thiele.
\newblock {Zippy: On-Demand Network Flooding}.
\newblock In {\em Proceedings of the $13^{th}$ International Conference on
  Embedded Networked Sensor Systems ({SenSys})}, pages 45--58. ACM, November
  2015.

\bibitem{suzuki15ewsn}
Makoto Suzuki, Chun-Hao Liao, Sotaro Ohara, Kyoichi Jinno, and Hiroyuki
  Morikawa.
\newblock {Wireless-Transparent Sensing}.
\newblock In {\em Proceedings of the $14^{th}$ International Conference on
  Embedded Wireless Systems and Networks ({EWSN})}, pages 66--67. Junction
  Publishing, February 2017.

\bibitem{sommer16channelhopping}
Philipp Sommer and Yvonne-Anne Pignolet.
\newblock {ependable Network Flooding Using Glossy with Channel-Hopping}.
\newblock In {\em Proceedings of the $17^{th}$ International Conference on
  Embedded Wireless Systems and Networks ({EWSN}), competition session}, page
  303. Junction Publishing, February 2016.

\bibitem{escobar2016redfixhop}
Antonio Escobar-Molero, Francisco~J Cruz, Javier Garcia-Jimenez, Jirka Klaue,
  and Angel Corona.
\newblock {RedFixHop with Channel Hopping: Reliable Ultra-low-latency Network
  Flooding}.
\newblock In {\em Proceedings of the $16^{th}$ International Conference on
  Design of Circuits and Integrated Systems ({DCIS})}, pages 1--4. IEEE,
  November 2016.

\bibitem{doddavenkatappa14p3}
Manjunath Doddavenkatappa and Mun~Choon Chan.
\newblock {P3: A Practical Packet Pipeline using Synchronous Transmissions for
  Wireless Sensor Networks}.
\newblock In {\em Proceedings of the $13^{th}$ International Conference on
  Information Processing in Sensor Networks ({IPSN})}, pages 203--214. IEEE,
  April 2014.

\bibitem{yuan15ripple}
Dingwen Yuan and Matthias Hollick.
\newblock {Ripple: High-throughput, Reliable and Energy-efficient Network
  Flooding in Wireless Sensor Networks}.
\newblock In {\em Proceedings of the $16^{th}$ International Symposium on A
  World of Wireless, Mobile and Multimedia Networks ({WoWMoM})}, pages 1--9.
  IEEE, June 2015.

\bibitem{mohammad18codecast}
Mobashir Mohammad and Mun~Choon Chan.
\newblock {Codecast: Supporting Data Driven In-Network Processing for Low-Power
  Wireless Sensor Networks}.
\newblock In {\em Proceedings of the $17^{th}$ International Conference on
  Information Processing in Sensor Networks ({IPSN})}, pages 72--83. IEEE,
  April 2018.

\bibitem{herrmann18mixer}
Carsten Herrmann, Fabian Mager, and Marco Zimmerling.
\newblock {Mixer: Efficient Many-to-All Broadcast in Dynamic Wireless Mesh
  Networks}.
\newblock In {\em Proceedings of the $16^{th}$ International Conference on
  Embedded Network Sensor Systems ({SenSys})}, pages 145--–158. ACM, November
  2018.

\bibitem{carlson13forwarder}
Doug Carlson, Marcus Chang, Andreas Terzis, Yin Chen, and Omprakash Gnawali.
\newblock {Forwarder Selection in Multi-transmitter Networks}.
\newblock In {\em Proceedings of the $18^{th}$ International Conference on
  Distributed Computing in Sensor Systems ({DCOSS})}, pages 1--10. IEEE, May
  2013.

\bibitem{brachmann16laneflood}
Martina Brachmann, Olaf Landsiedel, and Silvia Santini.
\newblock {Concurrent Transmissions for Communication Protocols in the Internet
  of Things}.
\newblock In {\em Proceedings of the $41^{st}$ International Conference on
  Local Computer Networks ({LCN})}, pages 406--414. IEEE, November 2016.

\bibitem{sarkar16sleepingbeauty}
Chayan Sarkar, R.~Venkatesha Prasad, Raj~Thilak Rajan, and Koen Langendoen.
\newblock {Sleeping Beauty: Efficient Communication for Node Scheduling}.
\newblock In {\em Proceedings of the $13^{th}$ International Conference on
  Mobile Ad Hoc and Sensor Systems ({MASS})}, pages 56--64. IEEE, October 2016.

\bibitem{rao_murphy_2016}
Vijay~S. Rao, Madhusudan Koppal, RangaRao Venkatesha~Prasad, T.V. Prabhakar,
  Chayan Sarkar, and Ignas Niemegeers.
\newblock {Murphy Loves CI: Unfolding and Improving Constructive Interference
  in WSNs}.
\newblock In {\em Proceedings of the $35^{th}$ International Conference on
  Computer Communications ({INFOCOM})}, pages 1--9. IEEE, April 2016.

\bibitem{wang_triggercast_2013}
Yin Wang, Yuan He, Dapeng Cheng, Yunhao Liu, and Xiang-yang Li.
\newblock {TriggerCast: Enabling Wireless Constructive Collisions}.
\newblock In {\em Proceedings of the $32^{th}$ International Conference on
  Computer Communications ({INFOCOM})}, pages 480--484. IEEE, April 2013.

\bibitem{wang_disco:_2015}
Yin Wang, Yunhao Liu, Yuan He, Xiang-Yang Li, and Dapeng Cheng.
\newblock {Disco: Improving Packet Delivery via Deliberate Synchronized
  Constructive Interference}.
\newblock {\em IEEE Transactions on Parallel and Distributed Systems},
  26(3):713--723, March 2015.

\end{thebibliography}

% Appendix
%\input{experiments_testbed}

\end{document}